\documentclass[]{aastex701}

\usepackage{amsmath,amsfonts,amssymb}

\usepackage{cleveref}

\usepackage{etoolbox}
\makeatletter
\pretocmd\@sect{\def\@currentcounter{#1}}{}{\fail}
\makeatother
\usepackage{booktabs}
\usepackage{multirow}

\newcommand{\R}{\mathbb{R}}

\begin{document}

   \title{A preliminary exploration of the effects of baseline length for the LIFE space mission}

\author[0000-0003-3992-342X]{Jonah T. Hansen}
\altaffiliation{Corresponding Author: johansen@phys.ethz.ch}
\affiliation{Institute for Particle Physics and Astrophysics, ETH Zürich, Wolfgang-Pauli-Strasse 27, Zürich 8093, Switzerland}
\email{johansen@phys.ethz.ch}

\author[0000-0002-2982-0390]{Thomas Birbacher}
\affiliation{Institute for Particle Physics and Astrophysics, ETH Zürich, Wolfgang-Pauli-Strasse 27, Zürich 8093, Switzerland}
\email{thomabir@phys.ethz.ch}

\author[0000-0002-5476-2663]{Felix A. Dannert}
\affiliation{Institute for Particle Physics and Astrophysics, ETH Zürich, Wolfgang-Pauli-Strasse 27, Zürich 8093, Switzerland}
\email{fdannert@phys.ethz.ch}

\author[0009-0000-7417-3201]{Philipp Huber}
\affiliation{Institute for Particle Physics and Astrophysics, ETH Zürich, Wolfgang-Pauli-Strasse 27, Zürich 8093, Switzerland}
\email{huberph@phys.ethz.ch}

\author[0000-0001-8450-3374]{Andrea Fortier}
\affiliation{Center for Space and Habitability, University of Bern, Gesellschaftsstrasse 6, Bern 3012, Switzerland}
\affiliation{Weltraumforschung und Planetologie, Physikalisches Institut, University of Bern, Gesellschaftsstrasse 6, Bern 3012, Switzerland}
\email{andrea.fortier@unibe.ch}

\author[0000-0001-9250-1547]{Adrian M. Glauser}
\affiliation{Institute for Particle Physics and Astrophysics, ETH Zürich, Wolfgang-Pauli-Strasse 27, Zürich 8093, Switzerland}
\email{glauser@phys.ethz.ch}

\author[0000-0003-2769-0438]{Jens Kammerer}
\affiliation{European Southern Observatory, Karl-Schwarzschild-Straße 2, Garching 85748, Germany}
\email{Jens.Kammerer@eso.org}

\author[0000-0002-2215-9413]{Romain Laugier}
\affiliation{Institute of Astronomy, KU Leuven, Celestijnenlaan 200D, Leuven 3001, Belgium}
\email{romain.laugier@kuleuven.be}

\author[0000-0001-8020-3884]{Lia Sartori}
\affiliation{Institute for Particle Physics and Astrophysics, ETH Zürich, Wolfgang-Pauli-Strasse 27, Zürich 8093, Switzerland}
\email{lias@phys.ethz.ch}

\author[0000-0003-3829-7412]{Sascha P. Quanz}
\affiliation{Institute for Particle Physics and Astrophysics, ETH Zürich, Wolfgang-Pauli-Strasse 27, Zürich 8093, Switzerland}
\affiliation{Department of Earth and Planetary Sciences, ETH Zürich, Sonneggstrasse 5, Zurich 8092, Switzerland}
\email{sascha.quanz@phys.ethz.ch}

  \begin{abstract}
   By aiming to find and characterise dozens of habitable exoplanets through the technique of nulling interferometry, the LIFE space mission will produce transformational science. One of the key parameters for such an interferometric mission is the nulling baseline length - the distance between nulled apertures, which past studies have assumed to be 10-100\,m. Advances in planet occurrence statistics and simulation tools allow us now to revisit this key assumption with significantly more detail, particularly with the intention to reduce the range of baselines considered due to mission implementation concerns. We utilise the LIFEsim mission simulator along with revised mathematical tools to identify whether the range of baselines could be reduced without significantly affecting planet yield and fringe tracking performance. Along the way, we also determine a new astrophysically motivated technique for choosing which baselines are optimal for a given science target. We find that indeed, LIFE could utilise a considerably shorter range of baselines, such as 25-80\,m, or even discrete baselines without much ($<$10\%) loss of performance. Nevertheless, careful trade-offs between performance and implementation simplification must be made, especially considering any spectral weighting that may be required by the scientific goals, and the potential loss of target-specific baseline optimisation.
    \end{abstract}


\section{Introduction}
\label{sect:intro}  

The LIFE (Large Interferometer For Exoplanets) space mission \citep{Quanz-2022-ID17,Glauser-2024-ID1} is \added{an ambitious observatory} with a science goal of finding and characterising between 30 and 50 terrestrial exoplanets around nearby solar-type stars. These characterisations \added{will assist in} searching for signs of life in these distant worlds through an analysis of spectral biosignatures \citep{Schwieterman-2018-ID60}. Such a goal is near-impossible for the current and near future fleet of ground and space based telescopes, requiring observations that are simultaneously high contrast, high sensitivity and high angular resolution to successfully record the spectra of these planets with sufficient signal. To overcome these hurdles, LIFE is designed as a space-based, formation-flying, mid-infrared (MIR) nulling interferometer operating between 4 and 18.5\,µm. 

Similar missions were proposed previously by ESA and NASA in the early 2000s, namely \textit{Darwin} \citep{Kaltenegger-2005-ID61,Cockell-2009-ID56} and TPF-I respectively \citep{Lawson-2008-ID62}, but were ultimately not adopted. Since then, significant progress has been made utilising planet population studies \citep{Kopparapu-2018-ID55,Bryson-2021-ID42} based on \textit{Kepler} statistics \citep{Borucki-2010-ID59} and the scientific utility of such a mission verified via numerous planet yield and spectral retrieval based studies \citep[e.g.][]{Quanz-2022-ID17,Kammerer-2022-ID41,Konrad-2024-ID64,Hansen-2025-ID51}. The mission reference architecture has nevertheless remained very similar to the convergent Emma X-array \& double Bracewell design proposed in the late 2000s \citep{Martin-2007-ID63}, consisting of four collector telescopes flying in a 6:1 rectangular formation that rotates about the target's line-of sight. The combiner lies out of plane at a distance of approximately 1\,km. More information on the reference architecture for the LIFE mission can be found in \cite{Glauser-2024-ID1}.

Within all these studies, some of the adopted parameters have remained unchallenged - primary among which is the \added{permitted} baseline length range of 10-100\,m. Rather, they have been specified as order-of-magnitude numbers that are roughly derived from nulling baseline limits of the earlier \textit{Darwin} (7-168\,m) \citep{Cockell-2009-ID56} and TPF-I (20-66\,m) \citep{Martin-2007-ID63} mission concepts. The current concept specifies that the nulling (i.e. shorter) baselines can range between 10 and 100\,m, with the longer imaging baselines thus ranging between 60 and 600\,m. For a given target, the interferometer is able to choose the baseline that is best optimised for obtaining signal from the planet and minimising noise from astrophysical sources. 

\added{The choice of} baseline can greatly influence the technical difficulty of the interferometric array's motions and stability. On the side of the maximum baseline, longer baselines result in the need for more complex metrology systems that can work over vast distances, and impede the ability to test systems on the ground. It also \added{increases the} array focal length (and hence a larger radius of curvature), as that distance is set from differential polarisation effects \citep{Defrere-2010}. This in turn leads to more stringent pointing requirements. 

The minimum baseline is even more critical, influencing margins for collisions, gravitational and electrostatic attraction, and thruster contamination among other effects. The current minimum baseline of 10\,m is likely already too short: estimates of the thermal requirements of a 3\,m collector mirror require a $6-8$\,m spacecraft including baffling (F. Konnerth and A. M. Glauser, priv. comm), leaving a tiny margin of a few meters between spacecraft. Hence justifying the increase of the minimum baseline without compromising the science is important to relax margins on the spacecraft formation.

In this study, we justify the baseline ranges for LIFE from a few various angles. Firstly, we examine a pseudo-analytical approach to optimising the baseline for a given target - whether it is for detection or characterisation. We then analyse how changing the baseline ranges affect the detection yields of LIFE, as well as which targets are most affected by constraining either the minimum or maximum baseline. We also consider the effects on fringe tracking, as well as explore the possibility of having only one or a few discrete baselines, and how this impacts the scientific performance as a trade-off against engineering complexity.

\section{The performance metric}

Before we start answering the questions listed above, it is worth reconsidering what metric we use to define performance. In this study, we only model astrophysical photon noise sources (of the type implemented by \textsc{LIFEsim} and similar studies \citep{Quanz-2022-ID17,Dannert-2022-ID11,Hansen-2022-ID6}). \added{As described succinctly in previous studies \citep{Dannert-2022-ID11,Hansen-2022-ID6}, the double Bracewell differential nulling technique, as foreseen to be used by LIFE, removes any coherent contribution of the noise sources to the measurement (i.e. parasitic signals), and so the contribution is reduced to photon shot noise (i.e. a DC offset).} We omit any effects of instrumental instabilities (discussed at length in \cite{Dannert-2025-ID49,Huber-2025-ID58,Rutten-2026}), as they are believed to not depend strongly on the length of the baseline but rather the baseline ratio, which we keep constant at the current 6:1 imaging to nulling baseline ratio \citep{Glauser-2024-ID1}. This choice has the consequence of us being able to use the signal-to-noise ratio (SNR) based on photon statistics, or equivalently the integration time needed to reach a given SNR, as a metric for performance as opposed to performing a full signal extraction with template matching \citep[as done in e.g.][]{Huber-2025-ID58}.


There are four sources of photon \added{flux} we consider here. First is the signal of the planet ($F_{p}$), and then three noise sources: residual starlight known as ``stellar leakage'' ($F_{s}$), light from the local zodiacal cloud ($F_{z}$), and light from dust around the target system known as exozodiacal light ($F_{e}$). Expressing these in terms of their flux for the $i$th wavelength channel we have:
\begin{equation}
    \text{SNR}_{\lambda(i)} = \frac{F_{p,i}}{\sqrt{F_{p,i}+F_{s,i}+F_{z,i}+F_{e,i}}}\sqrt{\pi D^2t\eta},
\end{equation}
where the latter factor is a common conversion to photon count: $t$ is the exposure time, $\eta$ is the system throughput and $D$ is the aperture diameter (a factor of four is removed due to having four collectors). A further assumption here is that the system throughput is wavelength independent, which while not strictly true, requires too many other assumptions on the instrument implementation. As with previous studies \citep{Dannert-2022-ID11,Hansen-2022-ID6}, we will take the total bulk-SNR as being the weighted SNR over each wavelength channel:
\begin{equation}
\label{eq:snr}
    \text{SNR}_\text{total} = \sqrt{\sum_i\text{SNR}_{\lambda(i)}^2}.
\end{equation}
We note that this last assumption, that noise is independent between wavelength channels, is another by-product of our omission of \added{explicit} instrumental noise; studies have shown that instrumental noise is in fact highly spectrally correlated, which hence require alternative treatments \citep{Huber-2025-ID58}. Discussions on how useful this bulk SNR metric is for planet characterisation can be found in \cref{sec:optimised_characterisation}.

\added{To account for the instrumental noise that is inevitably present, we take the same approach as previous LIFE studies \citep{Quanz-2022-ID17,Dannert-2022-ID11,Hansen-2022-ID6}), and claim a detection of a planet if we reach an SNR $\geq$ 7. This stems from the classic detection threshold of SNR $\geq$ 5 in the astrophysical photon noise limit, and then adding an equal portion of instrumentation noise. As instrumental instabilities primarily lead to an increase in the stellar leakage, the astrophysical and instrumental noise should scale in the same manner when dealing with brighter or fainter targets. \cite{Huber-2025-ID58} indicate that the SNR $\geq$ 7 criterion may be too conservative, finding that if LIFE's instrumental errors are properly calibrated and correlations accounted for, then the perturbed instrument is similar in performance to the unperturbed case. Nevertheless, we stick with the conservative estimate, which hence accounts for any residual noise left after the data is appropriately calibrated.  Work is ongoing to tie architectural trades to a proper treatment of instrumental noise.}

For the purposes of this work, we have aimed to optimise the calculation of each of the flux contributions. As such, we have produced pseudo-analytical expressions for each term rather than a numerical grid calculation as was done previously \citep{Dannert-2022-ID11,Hansen-2022-ID6}. This is described in \cref{app_sources}. We include some secondary effects previously neglected, namely that of limb darkening, which is further covered in \cref{app_limb_darkening}. 

Also absent from previous studies was an investigation into how the single-aperture field-of-view (FOV), driven by the use of a spatial filter, affects the measurement. Large collectors may result in desirable targets lying outside this FOV, thus resulting in a greater loss of signal compared to a smaller aperture. We include a short discussion on this effect in \cref{app_FOV}, as it also affects the procedure to optimise the baseline for particularly close targets. The conclusion is that, with the exception of the closest system $\alpha$ Cen, the FOV of apertures up to 5\,m in diameter do not greatly restrict the ability to observe the habitable zone of nearby stars. For the rest of this study we adopt a Gaussian FOV taper given as
\begin{equation}
\label{eq:fov}
    \rho(\theta,\lambda,D) = e^{-\frac{\pi^2}{4}\left(\frac{\theta D}{\lambda}\right)^2},
\end{equation}
where $\theta$ is the off-axis angle in radians and $\lambda$ the wavelength.

Finally, for the rest of this study we will focus on the length of the shorter ``nulling'' baseline, which we denote $B$. This is the baseline which creates the nulls in a double Bracewell architecture, and hence has \added{a direct} effect on performance. The longer imaging baseline is used for signal extraction and as stated earlier, can be calculated from our constant 6:1 baseline ratio. 

\section{The optimal baseline per target}
\label{sec:optimisation}
The first step in our analysis is to take a closer look at the method of optimising the baseline. In previous papers \citep{Dannert-2022-ID11,Hansen-2022-ID6,Kammerer-2022-ID41}, the baseline was optimised according to the following formula:
\begin{equation}
\label{eq:ref_wav}
    B_\text{opt} = \frac{0.59\lambda_\text{ref}}{\theta_p}
\end{equation}
where $\theta_p$ is the angular position of the desired optimisation (the location of the planet for characterisation, or the centre of the habitable zone for detection) in radians, the factor 0.59 derived from the location of the peak of the modulation efficiency curve $\xi(x)$ (\cref{eq:mod_eff}, \cref{app_sources}), and $\lambda_\text{ref}$ the so called ``reference wavelength''. This latter parameter corresponds to the wavelength for which a spectrally uniform source at that angular position has the highest throughput; this is less descriptive when discussing non-uniform spectral objects or a collection of sources at varying angular positions. This parameter also has the problematic by-product of having the same name as a different concept used by the LIFE science community, namely the wavelength used to scale the SNR of a retrieved spectrum. The value of $\lambda_\text{ref}$ used in previous works has generally been 15\,µm, found through a loose grid search and optimised for a given realisation of a synthetic planet population. 

Instead, we tackle this problem anew from first principles and verify it against yields. We will aim to find a description for the optimum baseline of the form 
\begin{equation}
    \label{eq:opt_B}
    B_\text{opt} = K\theta_p^{-1},
\end{equation}
where $K$ is denoted the baseline scaling factor, equal to $0.59\lambda_\text{ref}$. We separate the problem in two: optimising the signal while characterising a known planet, and optimising the number of detections when searching for unknown targets. 

\subsection{Characterisation}
\label{sec:optimised_characterisation}
For the characterisation of a known planet, if one wishes to optimise for the total SNR over the wavelength range of interest, we simply find $B$ that maximises \cref{eq:snr} for that target.
However, finding $K$ from \cref{eq:opt_B} is non-trivial, as this scaling parameter has a dependence on many parameters including the planet temperature ($T_p$), stellar effective luminosity ($L_\star$), stellar radius ($R_s$), distance ($d$), collector diameter ($D$), planet projected separation ($a_\text{proj}$), and exozodi-level ($z$). A brief discussion on how these effects affect the optimal baseline can be found in \cref{app_optimised}.

To factor in these effects, we performed a Monte-Carlo (MC) analysis over these variables ($T_p\in[150,600]$\,K, $L_\star\in[0.0005,6]\,L_\odot$, $d\in[1,30]$\,pc, $D\in[1,5]$\,m, $a_\text{proj}\in[0.01,1.75]$\,au, $z\in[1,10]$) and fitted a second-order multivariate polynomial to approximate the scaling factor $K$. \added{We disregard stars with a luminosity larger than that of any real world counterparts (adapted from the Gaia Nearby Star Catalogue, \cite{2021GaiaCollaboration}) at that given distance or closer, ignoring Alpha Centauri and Procyon as nearby special cases.} We factor in the $R_s$ dependence through interpolating the relationship between stellar temperature and radius on the main sequence tables of \cite{Pecaut-2013-ID47} to obtain the stellar luminosity of a given target. The resultant fit is of the following form (also used in the latter detection analysis):
\begin{equation}
    \label{eq:k}
    K = c_1T_p + c_2T_p^2+c_3a_\text{proj} + c_4a_\text{proj}^2+ c_5L_\star^{0.5} +c_6L_\star+c_7d + c_8d^2+c_9L_\star^{0.5}d+ c_{10}\theta_p^2+ c_{11}
\end{equation}
where $T_p$ is the planet equilibrium temperature, $a_\text{proj}$ is the projected physical separation in AU, $d$ is the distance in parsecs, $L_\star$ is the stellar luminosity in solar units and $\theta_p$ is the angular separation in radians. It should be noted that the fit did not depend strongly on $D$ or $z$ and so they were not included in the polynomial. It was also found that a piecewise approximation was needed, and so the sample was split into planets of temperatures greater and less than 350\,K, and between stellar types of M and FGK (a luminosity split at 0.08\,$L_\odot$). The parameter fits are in the first part of \cref{tab:opt_baseline_MC} in appendix \ref{app_coeffs}. Residuals for the coefficient fit are also found in appendix \ref{app_coeffs}, with a standard deviation of 1.7$\times 10^{-7}$\,m\,rad. We find that for characterising a known exoEarth twin ($T_p = 254$\,K) at 10\,pc, the scaling parameter is on the order of 7.7$\times 10^{-6}$\,m\,rad. This corresponds to $\lambda_\text{ref}\approx 13.0$\,µm, which is considerably shorter than that of 15\,µm.

Of course, for characterisation, it is also true that often not all parts of the spectrum should be weighted equally: it is often the case that certain lines or \added{portions} of the spectrum (perhaps short wavelengths) should be treated as more important than those at other wavelengths. This has been highlighted in a number of recent works \citep[see][]{Braam-2026-ID71,Fujii-2026-ID70}, whereby optimising for the bulk-SNR over the full wavelength range in-fact misses otherwise observable features if it were instead optimised for shorter wavelengths. 

To implement such wavelength dependent optimisation, weights need to be applied to each spectral channel in \cref{eq:snr} and the analysis rerun. However, as such weights are highly specific to the science question being asked, and as the LIFE science strategy is still being formulated, we do not address here how this will affect the needs of the baseline. Nevertheless, we do comment that this is one of the strengths of the flexible baseline approach - the baseline can be chosen based on the known planetary architecture and question being asked, to maximise the information for that specific target. Further studies that tackle the intersection between baseline optimisation and the retrieval of various biosignature gases would be highly fruitful in navigating the needs of the LIFE mission.

\subsection{Detection}
\label{sec:optimised_detection}
For detection, the question becomes more complicated, as one does not optimise for a single planet but rather the theoretical statistical population around the star. For LIFE, we are interested in maximising the detection yield of planets located within the habitable zone, defined as the region where liquid water may be present around a host star, and for this we take the main-sequence limits suggested by \cite{Kaltenegger-2017-ID48}. As a naive implementation, one would simply take the aforementioned \cref{eq:opt_B} for the characterisation case and optimise for the centre of the habitable zone. However, there are two effects that necessitate a revision of the scaling factor from that implementation.

First is the effect of orbital projection: planets cannot be assumed to be located at the projected angular distance formed by the semi-major axis and the distance to the system. Rather, they may be found at a closer projected distance due to being closer or further away from the star with respect to us as observers. In general, they may lie on any location on the surface of an ellipsoid with a given semi-major axis. For simplicity, and also driven by findings that Earth-like planets tend to have circular orbits \citep{Kipping-2025-ID43}, we will restrict our planetary orbits to zero eccentricity; meaning that all orbital realisations lie on the surface of a sphere. When projected into the plane of the sky, this means that for a given semi-major axis $a$, the probability distribution of projected separations is 
\begin{equation}
\label{eq:proj_orbit}
    P(a_\text{proj}) \propto \frac{a_\text{proj}}{\sqrt{a^2-a_\text{proj}^2}}, 0\leq a_\text{proj}<a.
\end{equation}
If we are restricting our detections to those planets inside the habitable zone, this leads to requiring longer baselines than otherwise predicted: these planets may be projected further in towards the star, and hence need longer baselines to \added{resolve} the signal for these targets. 

Secondly, there is the question of planet occurrence rates: from \textit{Kepler} \citep{Borucki-2010-ID59} and TESS \citep{Ricker-2015-ID67} statistics, it is expected that planets in the habitable zone tend to be more frequently distributed at smaller $a$. This is shown clearly from prior occurrence studies such as that of the \textsc{SAG13} report of \cite{Kopparapu-2018-ID55}, who model the planet distribution as proportional to an orbital period power law, $T^{\beta_\text{SAG13}-1}$; and \cite{Bryson-2021-ID42}, who model the frequency of planets with a stellar insolation of $I$ as proportional to $I^{\beta_\text{Bryson}}$. When converted to a distribution in $a$, these become:
\begin{align}
\label{eq:sma_kepler}
\begin{split}
    P(a) &\propto a^{(1.5\beta_\text{SAG13}-1)}; \quad HZ_\text{inner} < a < HZ_\text{outer} \quad[\text{SAG13}]\\
    &\propto a^{-(2\beta_\text{Bryson}+3)}; \quad HZ_\text{inner} < a < HZ_\text{outer} \quad[\text{Bryson}]        .
    \end{split}
\end{align}
We consider the same populations as in \cite{Kammerer-2022-ID41}, using the values of the \textsc{hab2min} and \textsc{hab2max} populations of $\beta_\text{Bryson}=-0.84$ and $-1.19$ respectively, along with the baseline \textsc{SAG13} population for planets with $R_p < 3.4\,R_\oplus$\citep{Kopparapu-2018-ID55} of $\beta_\text{SAG13} = 0.26$. All of these distributions prefer shorter period planets and \added{therefore} longer baselines; thus these two effects together result in the discrepancy between the previously used reference wavelength of 15\,µm and the optimal one found in the previous section. It should be noted that the power law dependence on $a$ is \added{essentially} identical between the \textsc{hab2max} and \textsc{SAG13} distributions, with $P(a) \propto a^{-0.61}$. On the other hand, due to its low completeness correction for long period planets, \textsc{hab2min} has a steeper power law. As the former two distributions agree with each other, and as they have the weaker dependence (and hence bias should these distributions be found to be incorrect), we will adopt the $P(a) \propto a^{-0.61}$ as the designated ``\textit{Kepler}'' distribution correction.

To then find the optimal baseline when considering these effects, instead of finding the peak of \cref{eq:snr} as a function of baseline, we instead find the baseline that maximises the integrated product of \cref{eq:snr} and \cref{eq:proj_orbit} over projected distance from the star, marginalising over the distribution of $a$ given in \cref{eq:sma_kepler}. We use a blackbody with the mean temperature across the habitable zone as the planet spectrum to optimise for, factoring in the above distributions of $a$ and $I$ to calculate the expected value. 

Similar to the characterisation case, we ran an MC analysis and fit a multivariate polynomial of the form of \cref{eq:k} to approximate the scaling factor. The fit and residuals are provided in appendix \ref{app_coeffs}, with a residual standard deviation of $5.9\times10^{-8}$\,m\,rad. \added{We find that the fit only depended on two parameters: the distance $d$ and the stellar luminosity $L_\star$}. To also observe the bias from assuming the \textit{Kepler} distribution in $a$, we ran another MC analysis with a flat distribution of $a$ across the habitable zone. This too is shown in table \cref{tab:opt_baseline_MC}, with a residual standard deviation of $4.4\times10^{-8}$\,m\,rad. For the \textit{Kepler} distribution, we find that the scaling factor for an Earth twin at 10\,pc corresponds is about 9.5$\times 10^{-6}$\,m\,rad, or $\lambda_\text{ref} = 16.1$\,µm; for the uniform distribution, these values are 8.6$\times 10^{-6}$\,m\,rad and $\lambda_\text{ref} = 14.7$\,µm. These values confirm the earlier suspicions as to why the previously used ``optimal baselines'' were larger than if naively predicted: orbital projection effects and a higher density of close in planets result in lengthening the required baseline when optimising for detections.

\section{Verification and simulation setup}
\label{sec:verification}
With these estimates of the scaling factor, we ran yield simulations using \textsc{LIFEsim} \citep{Dannert-2022-ID11} to compare and verify the number of detected planets with each optimisation method (legacy reference wavelength, \textit{Kepler} distribution and uniform distribution). They were run using the \textsc{LIFEsim} ``baseline'' scenario, with a number of modifications as described below. This setup is also used for the following sections that discuss the effects of baseline limits on \added{detection yield}.
\begin{enumerate}
    \item A revised catalogue of stellar targets was used, combining the set of primarily main-sequence FGK type stars within 50\,pc from the Habitable Worlds Observatory input catalogue \citep{Tuchow-2024-ID68} and M-dwarfs from the previous LIFE target catalogue \citep{Menti-2024-ID53}. Combining these allows for a more complete sample of stars within the solar neighbourhood, with more robust parameter estimation. For efficiency reasons, all stars of stellar type M were limited to being within 30\,pc. This stems from there being an already large population internal to 30\,pc; the difficulty of observing these targets already at this distance; and due to the lower scientific priority of these stars compared to those of earlier stellar types (as they may be less likely to have planetary atmospheres \citep{VanLooveren-2025-ID54}).
    \item Throughput was set at 0.15, with a quantum efficiency of 0.6. This stems from \added{revised} estimations of the instrument throughput budget utilising the concept described in \cite{Glauser-2024-ID1} and verified through experimental measurements in the Nulling Interferometer Cryogenic Experiment (NICE) \citep{Ranganathan-2024-ID9,Birbacher-Hansen-2026}. This sets the total efficiency ($\eta$) similar to those from assumed by the TPF-I and \textit{Darwin} mission concepts at about 9-10\% \citep{Lawson-2007-ID57,Cockell-2009-ID56}. \added{If the throughput is found to be lower, then this can be compensated by either a proportional increase in mission time or with larger collectors, though one must trade the latter carefully with the decreased FOV. Both of these can result in substantial cost increases, hence the criticality of ensuring high instrumental throughput.}
    \item Aperture diameter was set at 3.5\,m and spectral resolution set at 100. Previous studies have set the aperture to between 2 and 3.5\,m, with most using a much more conservative throughput of $<$ 0.05 \citep{Quanz-2022-ID17,Dannert-2022-ID11,Hansen-2022-ID6,Glauser-2024-ID1}. From current yield estimates utilising the Bryson \textsc{Hab2min} populations with exoEarth occurrence rates of 0.15-0.25,  and applying the above throughput assumption, the best estimate to achieve LIFE's primary mission goal of surveying approximately 50 terrestrial planets around solar-type stars is between 3 and 4\,m \citep{Dannert-2025-Thesis}.
    \item All nulling baseline limits were removed, to allow the baseline to be truly ``optimal''. The ratio was nevertheless restricted to 6:1. The effects of restricting the range will be analysed in the following sections.
    \item The FOV taper described in \cref{eq:fov} was applied to observations.
    \item For these simulations, we only deal with the detection phase, due to the aforementioned complexity with defining a successful characterisation. Search time was set at 2.5~years with a 1:5 duty cycle (i.e. 2 years of observation time). We do not include multiple visits, which while being necessary for a successful identification of an interesting target, will only serve to increase/decrease the absolute yield; not changing how the relative yield varies with baseline limits.
    \item \added{The default \textsc{LIFEsim} scheduler was used. This is an implementation of the greedy algorithm \citep{dantzig1957discrete}, whereby the amount of integration time spent on a given star is distributed based on the likelihood of detecting a planet around said star for a set integration time. For more information see \cite{Dannert-2025-Thesis}. This is roughly equivalent to the Altrustic Yield Optimiser (AYO) developed by \cite{Stark-2014} for direct imaging missions.}
    \item Rather than stellar types, three stellar populations were considered that have been conceived to align with key scientific goals:
    \begin{enumerate}
        \item Solar-type stars of types F0V-K5V, which are the primary targets for the LIFE space mission. LIFE aims to detect 30-50 planets around this population.
        \item Low-luminosity stars of types K6V-M3V. These stars are not yet fully convective and so may still exhibit the potential for habitability. LIFE aims to detect 15-25 planets around this population.
        \item Very low-luminosity stars of types M4V-M9V. These stars exhibit high magnetic activity and induce tidal locking in their habitable zone, making habitable planets unlikely. Nevertheless, these are important for contextual observations. LIFE aims to detect 10-20 planets around this population.
    \end{enumerate}
    For each of these samples, we aim to achieve the upper end of their detection ranges.
    \item Exozodiacal dust levels are drawn from the HOSTS statistical analysis, with a median of 3~zodi \citep{Ertel2020}. \added{The dust distribution is considered to be face-on, axisymmetric and optically thin; a discussion on these assumptions is found in \cref{app_sources}.}
    \item As described earlier, planets are determined to have been detected if they achieve an SNR of 7, stemming from a detection criterion of an SNR of 5, and then accounting for an additional equal weight contribution of instrumental noise on top of the simulated astrophysical noise. 
    \item Finally, we used similar planet populations as in the analysis from \cite{Kammerer-2022-ID41}, following the planet occurrence rates from the \textsc{hab2min} and \textsc{hab2max} models of \cite{Bryson-2021-ID42} for FGK stars and the occurrence rates from \cite{Dressing-2015-ID69} for M stars (both the baseline values and the pessimistic lower uncertainty values). \added{The populations were generated with the tool \textsc{PPop} by \cite{Kammerer-2018-ID74}}. These have been analysed over 100 universes\added{, and only consider single planet systems. A comment on multi-planet systems is found in \cref{app_sources}.} We also ran a comparison with the baseline \textsc{SAG13} planet model akin to that used in \cite{Quanz-2022-ID17} and described in \cite{Kopparapu-2018-ID55}. ``Habitable'' planets are defined as those with radii between 0.5 and 1.5$R_\oplus$ and lying within their host star's habitable zone.
\end{enumerate}

\added{The yields from this verification simulation are shown in \cref{app_coeffs}. Ultimately, we find that the yields from either the \textit{Kepler} and uniform baseline optimisations do not vary by more than 3\% compared to the legacy reference wavelength for any of the planet or stellar populations; far below the intrinsic uncertainty in the planet population itself. If anything, one could perhaps note a minute increase in yield when using the \textit{Kepler} optimisation for early-type stars, though this result is statistically insignificant. Nevertheless, this is in itself a useful result, verifying that the new model produces similar yields to that used in previous studies, and gives insight into where the 15\,µm value originate from.} We will use the \textit{Kepler} optimisation polynomial for the baseline scale factor in the upcoming analyses.

\section{Restricting the baseline range} 
\label{sec:range}
\begin{figure*}
    \centering
    \includegraphics[width=\linewidth]{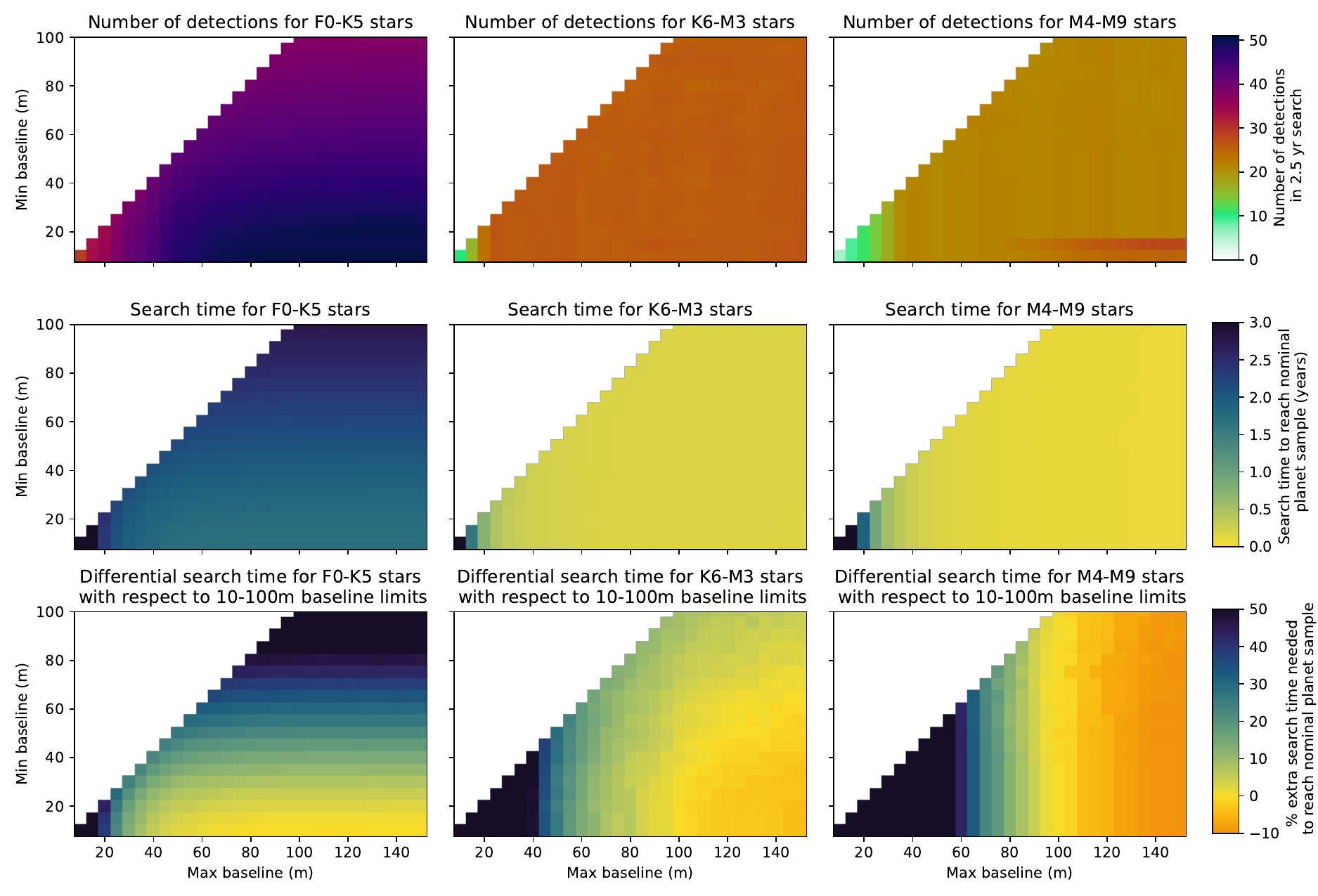}
    \caption{\textsc{LIFEsim} yield/mission time simulations for a range of baseline limits. The parameters for the simulation are detailed in the text, primarily \cref{sec:verification}. The columns denote the three stellar samples in decreasing order of scientific priority. \textit{Top row:} the raw yields for each baseline limit, given in planets detected. \textit{Middle row:} the total amount of search time needed to find the \added{nominal} planet sample for each stellar population (i.e. 50 \added{F0-K5} planets, 25 \added{K6-M3} planets, 20 \added{M4-M9} planets). \textit{Bottom row:} the percentage of extra search time needed to reach the same \added{nominal} planet sample compared to the reference 10-100\,m baseline case.\newline}
    \label{fig:range_big_plot}
\end{figure*}

We now reach the key investigation of this study, which is to determine a scientifically-motivated range of nulling baseline lengths. To achieve this, \textsc{LIFEsim} was used to run yield calculations on a grid of varying baseline ranges, using the same simulation parameters as detailed in \cref{sec:verification}. For the planet populations we use the \textsc{hab2min} model for FGK stars, and the baseline case from \cite{Dressing-2015-ID69} for M type stars; this gives us a mildly pessimistic assumption on the exoEarth occurrence rate. In the top row of \cref{fig:range_big_plot}, \added{we plot the number of planet detections within a 2.5 year search period (2 years of observation time) for the various stellar populations as a function of minimum and maximum nulling baseline length. In the middle row, we plot the mission time needed to detect (i.e. reach an SNR of 7 in a single visit) the nominal number of planets for each stellar sample (50, 25 and 20 planets for stellar types F0-K5, K6-M3 and M4-M9 respectively), and finally we plot the relative change in mission time needed compared to the 10-100\,m in the bottom row.}

It is clear that we achieve the required number of planets for many of these baseline configurations, with the low-luminosity samples in particular only exhibiting a substantial drop in detection ability for a maximum baseline less than 40\,m. We also see that the minimum baseline primarily sets the effectiveness of the \added{solar-type} yield, and the maximum baseline affects the low-luminosity star populations. For most configurations, the latter stars do not contribute substantially to the mission time; both populations can reach their target yield within a few months of search time. 




\begin{figure*}
    \centering
    \includegraphics[width=0.7\linewidth]{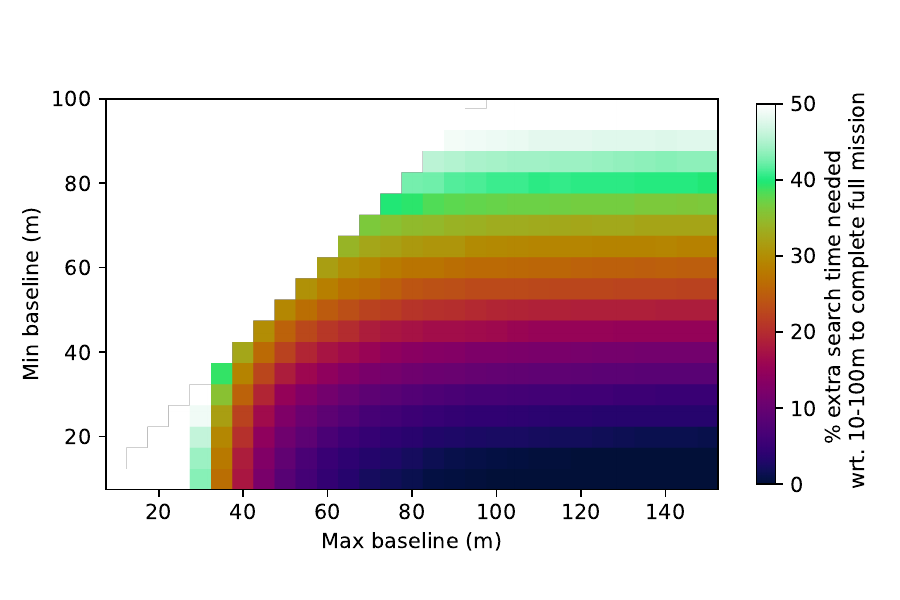}
    \caption{The percentage of extra search time needed to fulfil the targets of all three stellar populations (50, 25 and 20 planets for \added{stellar types F0-K5, K6-M3 and M4-M9} respectively) for a range of maximum and minimum baselines compared to the reference 10-100\,m case.}
    \label{fig:range_diff_time}
\end{figure*}

While the individual populations are useful for diagnosis, it is more important to look at the collective whole. In \cref{fig:range_diff_time}, we plot the relative mission time needed to accomplish the detection of all three planet samples compared to the 10-100\,m case. We see some clear trends, with setting a maximum baseline beyond 80\,m not providing much benefit; even a maximum around 50\,m does not lead to \added{a substantial} increase in mission time. This stems from most of the mission being spent detecting solar-type stars, which do not require long baselines. However, we must emphasise here that this analysis focuses on the detection of planets, and depending on the characterisation needs of LIFE, having the ability to probe atmospheres with longer baselines could be greatly advantageous. 

At least in the purview of detection, minimum baseline is a much more important factor, due to it's influence on the high priority solar-type targets. While having the smallest minimum baseline is ideal, we can see that increasing the minimum length only has a modest degradation in performance. Applying a rather arbitrary 5\% maximum mission time extension, the minimum baseline could be extended to 25\,m without much loss of performance. For the rest of this study, we will adopt a 25-80\,m range to explore other effects on the performance of the mission.

\begin{figure*}
    \centering
    \includegraphics[width=0.9\linewidth]{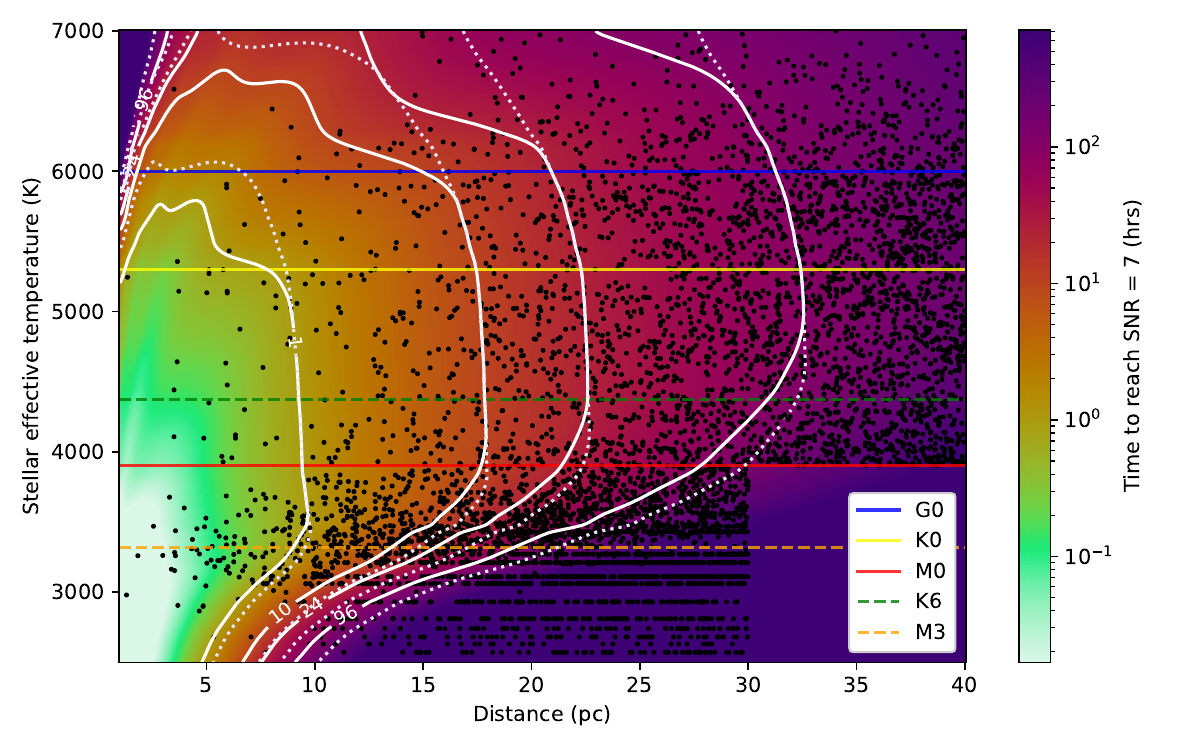}
    \caption{Integration time required for an Earth-twin around a given star to reach an SNR of 7, plotted as a function of stellar parameters. The baseline limits are set at 25-80\,m. The LIFE stellar sample is over-plotted in black, and integration time contours are over-plotted in white. Dotted contours show the same integration time contours, but for the 10-100\,m case for comparative purposes. Solid horizontal lines show the boundaries between stellar types of F, G, K and M; and dashed horizontal lines show the boundaries between LIFE's target stellar populations.}
    \label{fig:range_perf_time}
\end{figure*}

Now, the yield analysis focussed on the statistics of detection, rather than precisely which populations of stars are influenced by restricting the baseline limits. As such, we ran alternative simulations (using the mathematical tools described in appendix \ref{app_sources}) to calculate the amount of time needed to detect an exoEarth twin around a grid of stellar parameters for different baseline limits. We plot one such case in \cref{fig:range_perf_time}, with the colour map and solid contours showing the mission time needed for the 25-80\,m array, and the dotted contours comparing that to the 10-100\,m case. Overplotted is the current LIFE/HWO stellar catalogue, previously described in \cref{sec:verification}, and horizontal lines corresponding to the boundaries of various stellar types and populations.

\added{Here, we can see that with observations of $<$\,24 hours, we are sensitive to habitable exoplanets around all Solar-type stars out to 22\,pc}, and a substantial fraction of low-luminosity stars out to 15\,pc. Nearby early type stars suffer from having a baseline that is too long, although this effect is weaker in magnitude for G-dwarfs. The hottest F-dwarfs do have a substantial reduction in performance, but this only occurs at distances beyond 10-15\,pc. 

The very late M-dwarfs suffer both due to their limited luminosity and their inability to be optimised, despite the more favourable contrast. Here, much longer baselines would be needed in order to separate the planet flux from the star in a more efficient manner. This problem goes further when we consider signal extraction: if the baseline is too short, the planet may not encounter sufficient modulation to extract the signal, especially in the presence of instrumental noise. For this reason, only the very closest late-type stars may be accessible for LIFE unless the maximum baseline is substantially increased.


Overall, it appears that LIFE could reduce its nulling baseline range without much of an impact in performance; certainly with regards to its main targets of G and early K-type stars. However, this comes with the caveat that limiting the baselines also limits the ability for adaptation, and can substantially hinder the ability to analyse M-dwarfs. This corroborates the finding of \cite{Rutten-2026}, whereby one of LIFE's major advantages is its ability to analyse these cooler stars with flexible baselines. Of course, if the scientific priority of these targets is deemed low enough, and the mission implementation with these restrictions is shown to be significantly easier, then such a trade may be deemed acceptable. Taking this piece of logic to its conclusion leads us to a similar topic discussed in the aforementioned study: exploring discrete baselines. What if instead of being able to set a baseline anywhere within a range, LIFE could only choose out of a handful of well-defined baselines? 

\section{Discrete baselines}

The logic behind this investigation lies in the potential avenues to simplify some of the mission architecture. For example, if a single baseline length could suffice, then the formation-flying aspect of the mission could be removed entirely, or elements such as tethers considered. Alternatively, having only a few baselines could greatly reduce complexity in the optical train. One of the current issues surrounding the reference design is the large amount of astigmatism obtained by having single spherical collector mirrors that can change their angle with respect to the combiner (due to flexible baselines). This thus requires a high-stroke deformable mirror to correct such aberrations. If a subset of baselines are instead considered, one could imagine having a discrete set of correction optics for each baseline, alleviating the requirements on a deformable mirror and also allowing better optimisation for the optical train.

\begin{figure*}
    \centering
    \includegraphics[width=\linewidth]{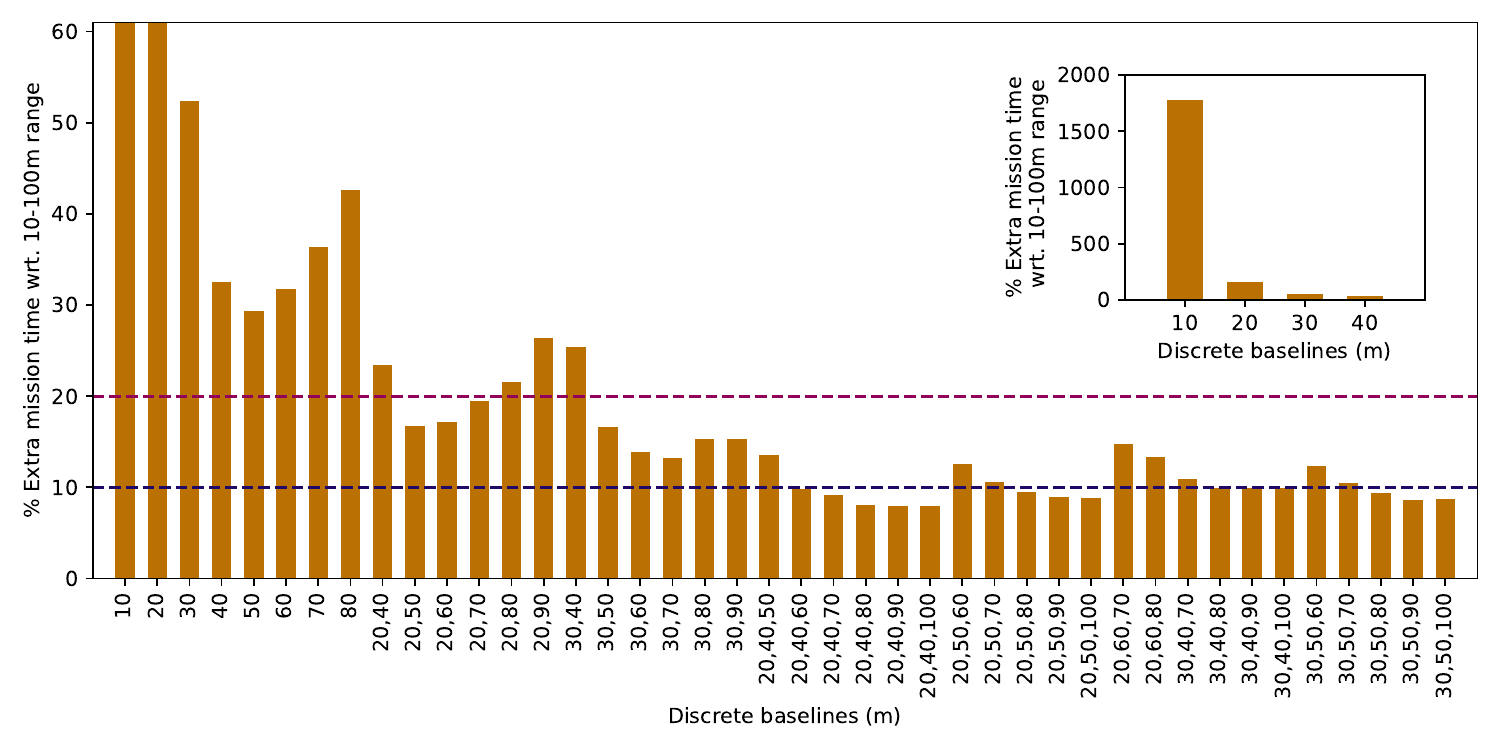}
    \caption{The percentage of extra mission time that is needed compared to the flexible 10-100\,m baseline array to detect the \added{total} number of planets specified in \cref{sec:verification} (50, 25 and 20 planets for \added{stellar types F0-K5, K6-M3 and M4-M9} respectively), plotted for a variety of discrete baseline cases. Dashed contour lines at 10\% and 20\% are also displayed. The plot inset uses a much larger y-axis scaling to demonstrate the impact of short single baselines.}
    \label{fig:discrete_yield}
\end{figure*}

To investigate this, \textsc{LIFEsim} yield simulations akin to the study in the previous section were undertaken. All combinations of 1, 2 or 3 baselines with intervals of 10\,m were compared to the reference flexible baseline case of 10-100\,m. When choosing between multiple baselines, we select the baseline that is closest to the optimal. For some pairs and triplets this may not be the correct selection due to the non-linearity of the baseline performance curve (see \cref{app_optimised}), but regardless allows the simulations to run in a reasonable timeframe. As such, the results presented are marginally conservative in performance. A subset of these combinations, notably the ones with the smallest increase in mission time, is displayed in \cref{fig:discrete_yield}. 

Having a single baseline of 10\,m is incredibly problematic for the mission, resulting in a staggering 18 fold increase in mission time. Beyond the short, single baselines, however, many of the other discrete cases do not perform badly. A single 50\,m baseline has a 30\% mission time increase; a 30 and 70\,m baseline pair has a 15\% increase; and a triplet of 30, 50 and 90\,m has a meagre 8\% increase in mission time. We currently discount the 20\,m triplets despite their limited performance degradation as they do not provide enough of a benefit over the 30\,m variants, which are much more favourable in terms of satellite formation margin.

\begin{figure*}
    \centering
    \includegraphics[width=0.8\linewidth]{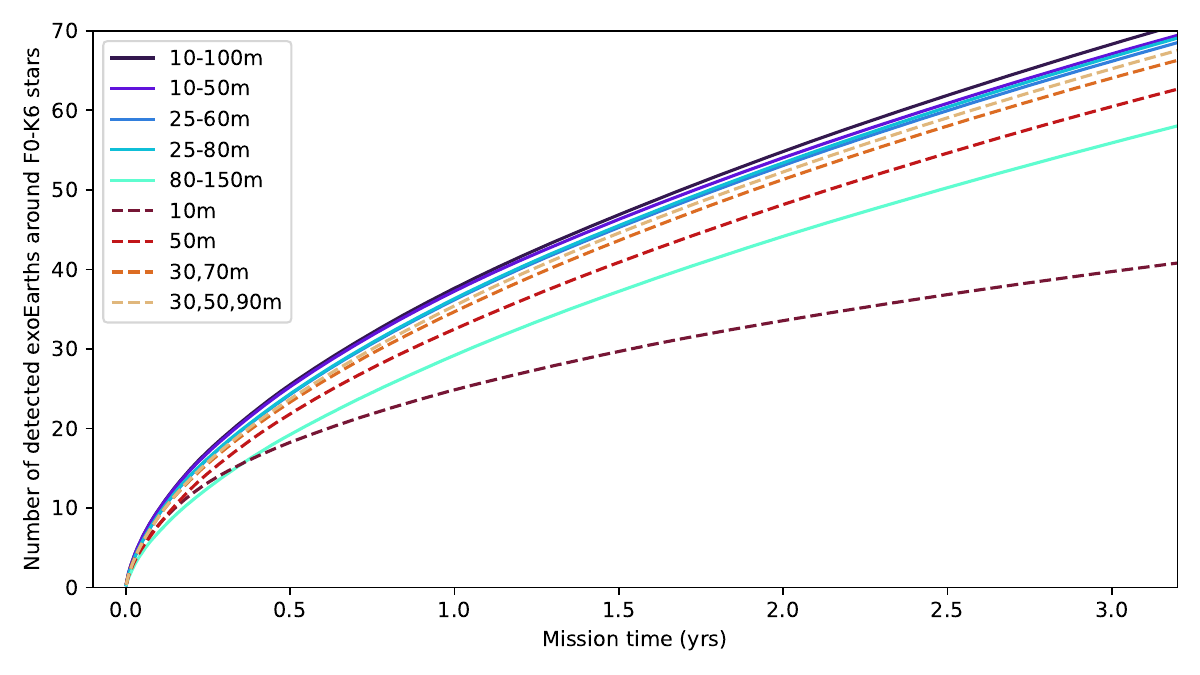}
    \caption{\added{Detection curve for a selection of baseline architectures, highlighting the cumulative number of exoEarth planet detections around solar-type stars as the mission unfolds. Baseline ranges are plotted as continuous lines, and discrete baselines as dashes.}}
    \label{fig:det_curve}
\end{figure*}

\added{We can see a different perspective of the same effect in \cref{fig:det_curve}, which plots the number of exoEarth detections around solar-type stars (F0-K6) as the mission progresses. We can clearly see that the derivative of the single baseline curves (especially 10\,m) is smaller than that of the ranges; that is, the number of planets we are able to detect plateaus faster. We can also see the conclusions drawn in the previous section: for solar-type stars the minimum baseline is critical, and that decreasing the range to 25-80\,m or even 25-60\,m does not lead to a big change in detection rates.}

\begin{figure*}
    \centering
    \includegraphics[width=\linewidth]{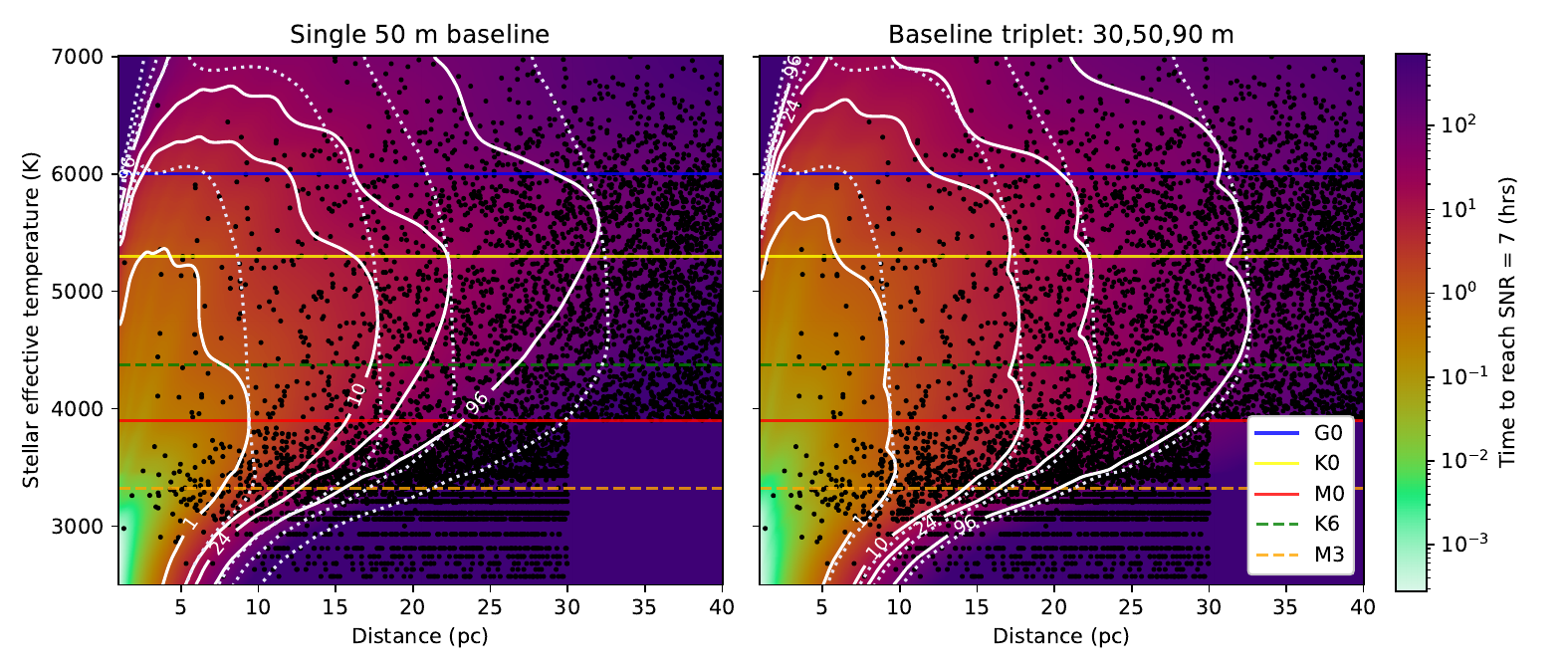}
    \caption{Integration time required for an Earth-twin around a given star to reach an SNR of 7, plotted as a function of stellar parameters, for two different discrete baseline configurations. The LIFE stellar sample is over-plotted in black, and integration time contours are over-plotted in white. Dotted contours show the same integration time contours, but for the 10-100\,m baseline range case for comparative purposes. Solid horizontal lines show the boundaries between stellar types of F, G, K and M; and dashed horizontal lines show the boundaries between LIFE's target stellar populations. \textit{Left:} Single 50\,m baseline; \textit{Right:} baseline triplet of 30, 50 and 90\,m.}
    \label{fig:discrete_perf}
\end{figure*}

Finally, we show a similar stellar parameter plot as in \cref{fig:range_perf_time}, which plots the extra observing time needed by a discrete array compared to a 10-100\,m flexible array for observing Earth twins around various stars. It's clear that each baseline adds a curve in stellar parameter space whereby the baseline remains optimised, and then the \added{performance degrades} as one deviates from that curve. G and K-type stars remain very well optimised for the baseline triplet case, though early-type stars suffer for the single baseline. For targets at longer distances and especially for low-luminosity targets, this becomes very problematic; the single 50\,m baseline loses a significant fraction of late K and M-dwarfs outside of the closest 10s of targets.  

Now, the results from this analysis indicate that indeed, from a statistical performance perspective, discrete baselines are a viable option for LIFE's architecture, whether that be through a connected structure or the ability to better correct aberrations in a formation flying setup. Nevertheless, we emphasise that there is still substantial benefits in keeping the flexible array architecture, namely through the ability to adapt baselines on-the-fly. One can imagine a few scenarios where this is advantageous: perhaps there are multiple planets in a given system that prevent the nominal optimal baseline from obtaining a significant SNR from the target planet \citep{Cacaj-2025-ID73,Rutten-2026}. Or, even more likely, the initial spectrum of the planet may show some unusual features that warrant further detailed investigation; such features then need to be optimised which can only be done with a flexible array architecture \citep{Braam-2026-ID71}. The trade-off as to whether these benefits, substantial though they are, overcome the benefits of architecture simplification due to discrete baselines will have to be made in the coming years.

\section{Fringe tracking}

Along with the scientific performance, there is another effect that we must consider, that of fringe tracking. The fringe tracker's role in the interferometer is to determine the precise, ground truth of the optical path length difference (OPD) through analysis of the stellar fringe pattern, likely at a similar albeit separate wavelength to that of the science. The ability to obtain sufficient \added{measurement precision} is dependent on the fringe contrast, or visibility, which itself is also dependent on baseline. Longer baselines result in a more resolved star, which can in turn lead to a decrease in fringe visibility and hence more noise on the OPD measurement. As such, we now analyse whether fringe tracking may impose limits to our desired baselines. 

To begin with, we require the visibility function, which is described by the Hankel transform of the stellar intensity distribution. We adopt a limb-darkening law as described in \cref{app_sources} and \cref{app_limb_darkening}, given by:
\begin{align}
\label{eq:limb_darkening}
    LD(\theta) &= 1-a_1(1-\mu)-a_2(1-\mu)^2,\quad
    \mu = \sqrt{1-(2\theta/\delta_s)^2},\quad\theta\leq\delta_s/2,
\end{align}
and from \cite{HanburyBrown-1974-ID35} we can formulate the visibility as:
\begin{align}
    \label{eq:visibility}
    V(x) &= \frac{1}{k_0}\left(\frac{k_1J_1(x)}{x}+\frac{k_2J_\frac{3}{2}(x)}{x^\frac{3}{2}}+\frac{k_3J_2(x)}{x^2}\right), \quad 
    x= \frac{\pi B\delta_s}{\lambda},
\end{align}
where $J_\nu$ are Bessel functions of the first kind, $k_n$ are the same coefficients linked to the limb darkening law as listed in \cref{eq:limb_params} and $\delta_s$ is the angular diameter of the star. 

Now, for our fringe tracker we assume that we encode the phase in a similar way to the GRAVITY photometric chip \citep{Benisty-2009-ID33}, whereby we do a redundant pair-wise static ABCD combination for each of the six distinct baselines. There are alternative combination schemes, such as an all-in-one combiner \citep{Anugu-2020-ID38,Lopez-2022-ID39}, or utilising a less splits via a tricoupler \citep{Hansen-2022-ID32}, but these trades are beyond the scope of this work. The phase estimation variance for such a combination scheme over a single baseline can be described following \cite{Blind-2011-ID34}:
\begin{align}
    \sigma_{\phi,\text{pair}}^2 \approx 2\left(\frac{4 \sigma_r^2+N}{N^2V^2}\right),
\end{align}
where $\sigma_r$ is the read noise and $N$ is the total number of photons obtained for a pairwise interferometric measurement (that is, from two apertures). We have made the assumption here that the piston has minimal variance during an integration; we do not have an atmosphere to contribute seeing effects and laboratory control measurements (1-10\,nm RMS from \cite{Birbacher-Hansen-2026}) indicate that instrumental effects will not cause a deviation of more than 1\%.

We also assume here that the fringe tracker will work in the Johnson K-band, spanning 2.2 $\pm 0.15$\,µm. The uncertainty in piston between each pair of telescopes needs to be on the order of 1\,nm, which corresponds to a phase variance of 10\,mrad; this value will be refined later. The variance is minimised when $V=1$ and so for a lower bound estimate on $N$, assuming a detector read noise of $<$1e-/DIT (e.g. a Hg-Cd-Te SAPHIRA detector \citep{Finger-2016-ID36}), we find that $N \approx 100000$\,ph. Clearly this is not a photon-starved regime and much larger than the readout noise component; as such, we neglect read noise moving forward. 

For a redundant array of $n_T$ telescopes with $0.5n_T(n_T-1)$ baselines, we have the flux of each telescope being distributed over ($n_T-1$) baselines. This thus increases the variance by that same factor. Also, in such a redundant scheme, we have 0.5$n_T$ measurements per differential piston. Hence the phase variance becomes:
\begin{align}
    \sigma_{\phi}^2 \approx 4\frac{(n_T-1)}{n_T}\left(\frac{1}{NV^2}\right) = \frac{3}{NV^2}.
\end{align}

So, converting to a differential piston measurement and expanding $N$ to integrate over the fringe tracker spectral bandpass gives:
\begin{align}
\label{eq:FT_sigx}
    \sigma_x &= \frac{\lambda}{2\pi}\sigma_\phi\approx\frac{\sqrt{6}}{\pi^2 D\delta_s\sqrt{t\eta_{FT}} }\int_\mathbf{\Delta\lambda}\frac{\lambda}{\sqrt{B_\lambda(T,\lambda)} V(B,\lambda,\delta_s)}d\lambda,
\end{align}
where $\eta_{FT}$ is the throughput up to the fringe tracker; $t$ is the exposure time for one set of piston measurements; and $B_\lambda(T,\lambda)$ is the Planck function. The limb darkening parameters used in the estimation of the visibility are calculated in \cref{app_limb_darkening}, yielding $k_0=0.465$, $k_1 = 0.654$, $k_2 = 0.616$ and $k_3 = -0.54$, for a maximum limb-darkened squared visibility of $V^2 = 0.83$. \added{We assume a fringe tracker throughput, including detector quantum efficiency, of 15\%; slightly higher than that of the science channel.}

\begin{figure*}
    \centering
    \includegraphics[width=0.9\linewidth]{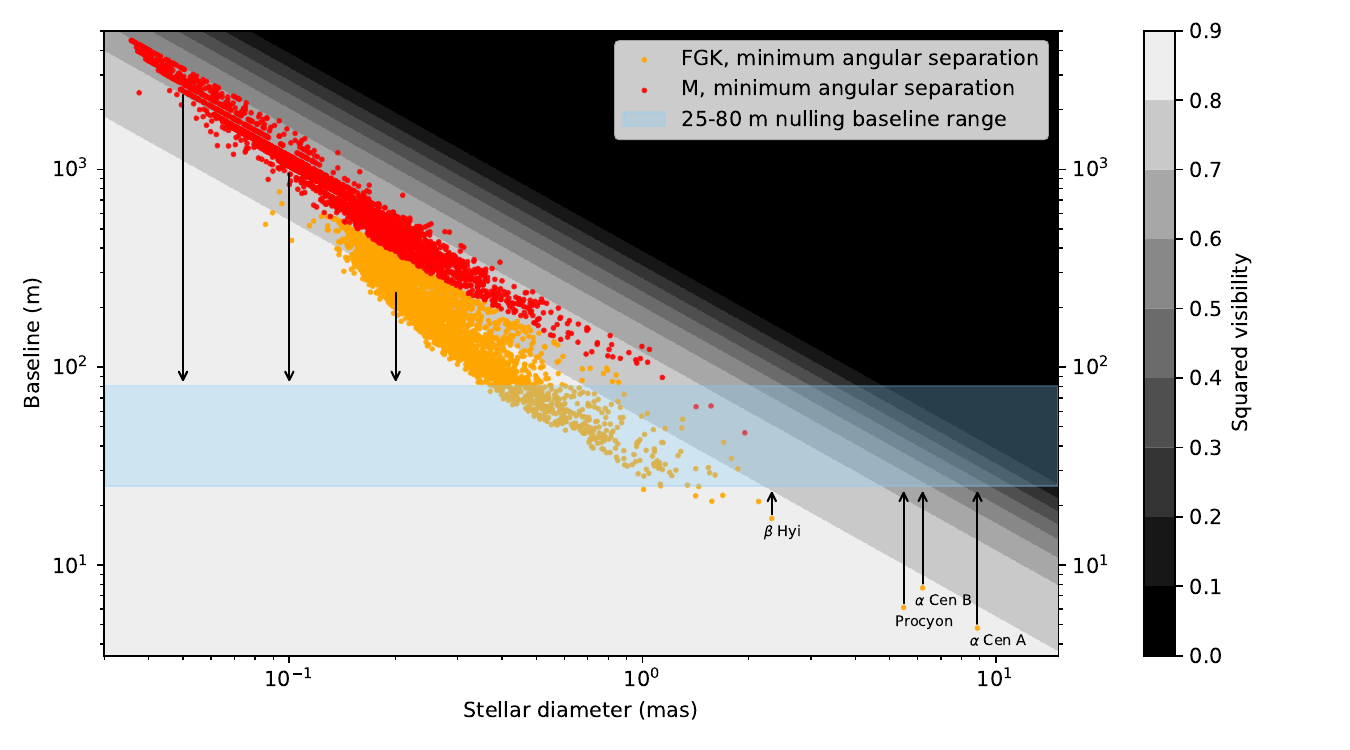}
    \caption{Squared visibility as a function of stellar angular diameter and baseline. Over-plotted is the LIFE stellar catalogue, whereby we choose the nulling baseline to correspond to optimising against a planet at half the inner edge of the habitable zone. More than 95\% of potential planets should lie at angular separations larger than this. The blue shaded region shows the 25-80\,m nulling baseline range as discussed in the previous sections. Arrows indicate that under a nulling baseline range restriction, stars that lie outside the shaded range are moved to the closest point inside the shaded region, which for short baselines may decrease visibility. The worst affected stars in this manner are labelled.}
    \label{fig:FT_vis_plot}
\end{figure*}

First, we ask the question whether resolving the stellar disk will ever be a problem for the fringe tracker. To give the worst-case scenario, we consider planets at the smallest angular separation that hence require the longest baselines. Accounting for orbital projection effects, approximately 95\% of all habitable zone planets should lie at projected separations larger than half the inner edge of the habitable zone. 
In \cref{fig:FT_vis_plot}, we plot how the visibility changes as a function of stellar diameter and baseline, and overplot LIFE's stellar targets assuming we are characterising a planet at that minimum angular separation, allowing the array to \added{freely} optimise the nulling baseline. The horizontal shaded region shows the range of nulling baselines we proposed in the earlier section. 

When we impose a baseline range restriction, we force stars that desire a larger baseline to be observed with shorter ones; this hence acts to increase the visibility of the stellar fringes. On the other hand, for stars that are forced to have longer baselines than optimal, we may end up further resolving the star and reducing the visibility. Fortunately, we have very few stars whereby this is substantially the case, namely $\alpha$ Cen A and B, Procyon, and $\beta$ Hyi; these are named in the figure. 

What we can draw from this plot is that visibility will not be a problem for the fringe tracker when considering the nulling baselines. For almost all stars and planetary configurations, we can expect a visibility drop of less than 10\%, highlighting that LIFE's nulling baselines are simply too short to resolve most of these main sequence stars. It is worth noting that fringe tracker visibility should only further increase for larger planet separations. The two notably resolved stars, that of Procyon and the $\alpha$ Cen system, still do not have a substantial visibility reduction, and their relatively high flux should counteract any visibility loss.

We must briefly cover the other sets of baseline lengths, namely the imaging baselines and the crossed baselines. These are substantially longer than the nulling baselines, both being approximately six times larger. In \cref{fig:FT_vis_plot}, this has the effect of shifting all points upwards on the figure, and hence has a large impact on the visibility as the star is now resolved. Nevertheless, this is not much of a concern as the phase requirement on the cross combiner is \added{considerably} less strict \citep{Hansen-2023-ID10}. As a first-order approximation, we assume that the requirement on the imaging baselines is an order of magnitude less strict (i.e. $>$10\,nm RMS). From \cref{eq:FT_sigx}, this results in a allowing an equivalent squared visibility 100 fold less, which is well within the visibility drop for almost all targets on these longer baselines. The aforementioned named targets will suffer though; $\alpha$ Cen A has a 200 fold drop in squared visibility, though again, the high flux of these nearby targets outweighs the relative drop in visibility.

Now, the key parameter to identify the fringe tracker's performance is the temporal bandwidth, or how fast we are able to run the system while retaining the required OPD residuals. The faster we can run the fringe tracker, the looser the requirements on other metrology measurements that do not provide the same level of ``ground truth''. The bandwidth can be calculated using \cref{eq:FT_sigx} (with the Nyquist-sampled bandwidth $f = 1/2t$), or equivalently using a star's K-band magnitude ($m_K$):
\begin{equation}
\label{eq:mag}
    \log_{10} f \approx \log_{10}\left[\frac{F_0\eta_{FT}\pi^3D^2V^2\sigma_x^2}{3\lambda_K^2}\right]-0.4m_K,
\end{equation}
where $F_0$ is the zero-point flux for the K-band and $\lambda_K$ is the central wavelength of the bandpass.

The fringe tracking residual $\sigma_x$ requirement is derived in an equivalent way to \cite{Birbacher-Hansen-2026}, where it is defined as the equally weighted error allowed such that the equivalent null depth $N_d$ does not lead to a degradation in the SNR of more than 10\%. We relate the residual and null depth via
\begin{equation}
    N_d \approx \frac{1}{2}\left(\frac{2\pi\sigma_x}{\lambda}\right)^2
\end{equation}
and we choose to use the shortest wavelength of 4\,µm as it sets the tightest requirement on stellar leakage. The stellar leakage for a given null depth is given in \cref{eq:flux_s}. 

\begin{figure*}
\centering
    \includegraphics[width=\linewidth]{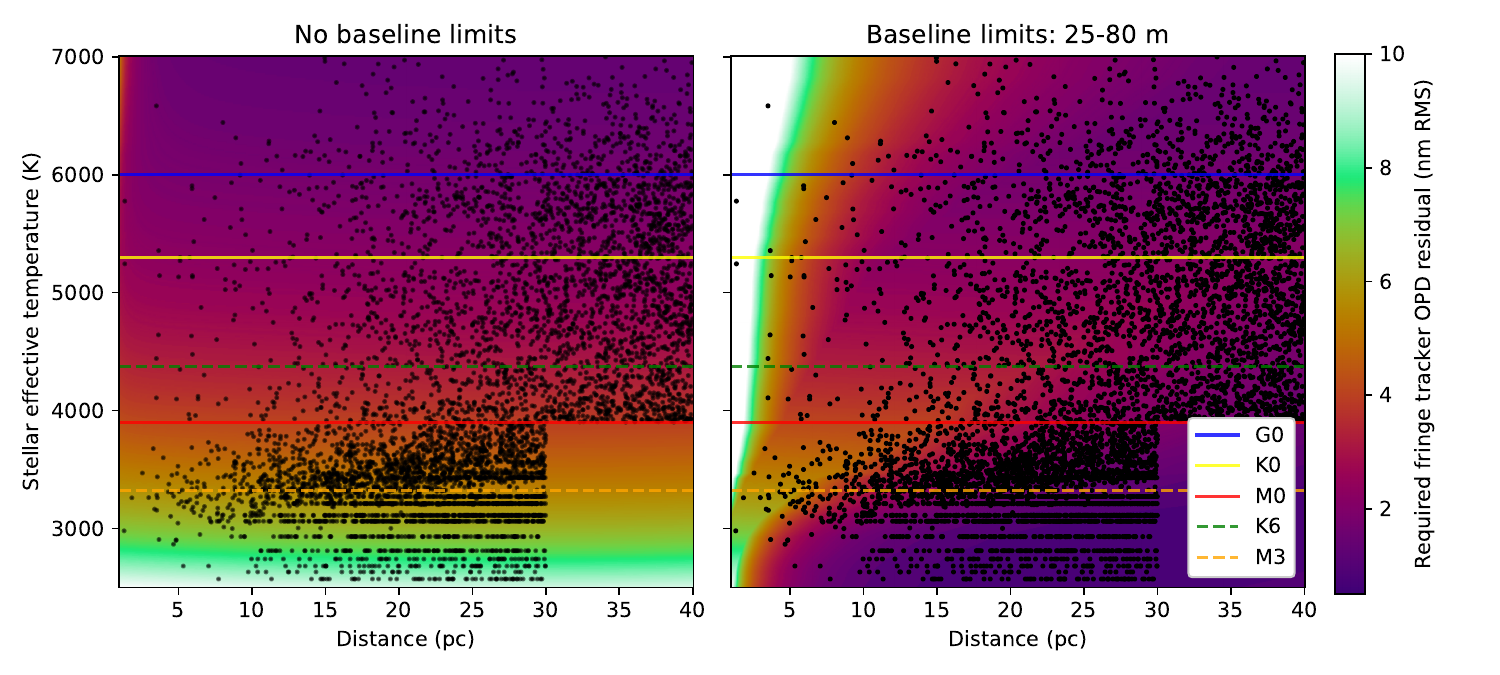}
    \caption{Requirements on the fringe tracking residuals for a grid of stellar parameters (distance and effective temperature). The LIFE stellar catalogue within 40\,pc is over-plotted in black. Solid horizontal lines show the boundaries between stellar types of F, G, K and M; and dashed horizontal lines show the boundaries between LIFE's target stellar populations. \textit{Left:} no restrictions on the baseline. \textit{Right:} baselines restricted between 25 and 80\,m.}
    \label{fig:FT_requirement}
\end{figure*}

The fringe tracker OPD requirement is shown as a function of stellar parameters in \cref{fig:FT_requirement}, both for the case without baseline limits, and for the 25-80\,m case that was proposed in \cref{sec:range}. Note that we further halve the requirement from what was calculated using the formula above, as the total residual cannot purely come from uncertainty in the measurement. We see that in the unrestricted baseline case, the residuals mostly scale with effective temperature, with cooler, fainter M-dwarfs not requiring as strict of an OPD control as the solar-type stars. 

Once the baselines are restricted however, the situation changes with M-dwarfs having much stricter requirements and nearby solar-type stars being looser. This effect is counter-intuitive. For the nearby solar-type stars, the baselines are too long, which results in a large increase in geometric leakage whereby the null is not \textit{broad} enough. This type of stellar leakage is not fixed through having a \textit{deeper} null and so null depth is no longer a limiting factor, easing the OPD requirements. 

On the other-hand, for the M-dwarfs, the baselines are far too short. In this regime, the ratio between the planet signal and the stellar leakage is much closer, and thus the SNR is more sensitive to null depth despite the more favourable contrast. This latter effect is very much a by-product of our definition of the OPD residual requirement, such that instrumental perturbations cannot degrade the SNR by more than 10\%. There is merit to the discussion of whether this is a viable metric to use; however, until a more robust instrument model is implemented \citep[the first efforts of which are found in ][]{Dannert-2025-ID49,Huber-2025-ID58}, this allows us to obtain an order-of-magnitude estimate. We also reiterate that observing planets around these M-dwarfs with short baselines will not yield substantial modulation and will be challenging for signal extraction beyond fringe tracking. The take-home result, nevertheless, is that the OPD residual requirement for most stars is on the order of 1-5\,nm RMS.


\begin{figure*}
    \centering
    \includegraphics[width=0.85\linewidth]{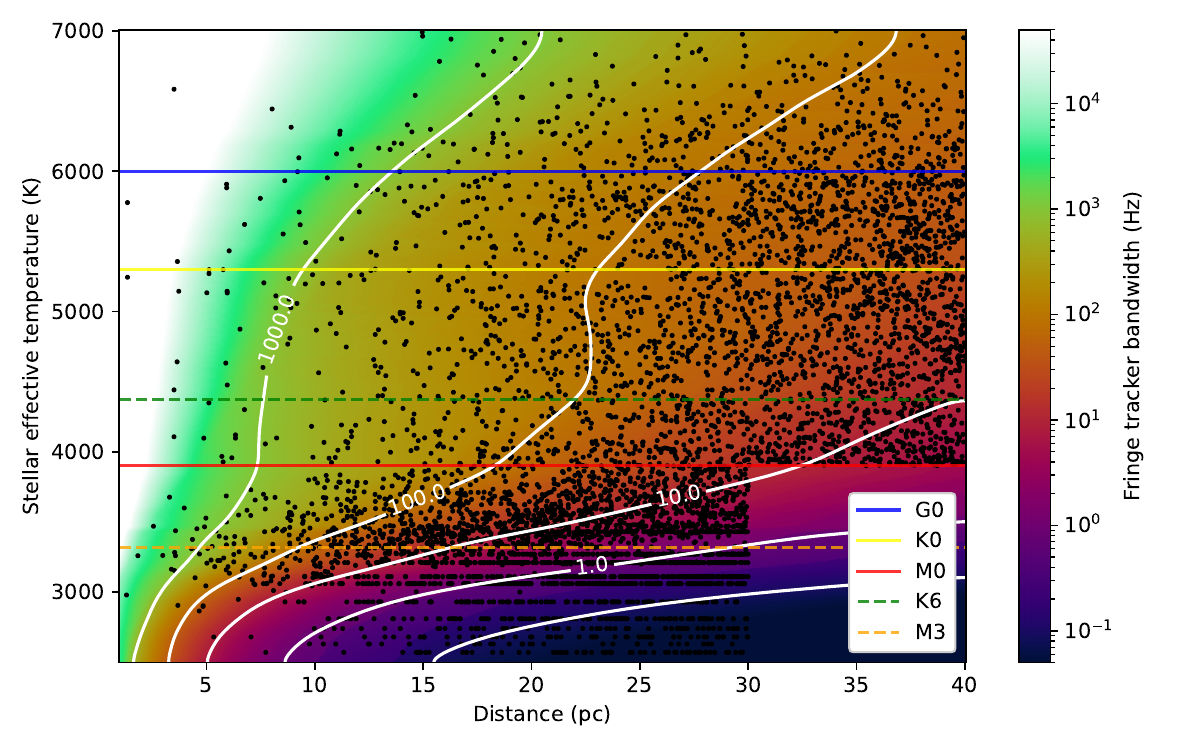}
    \caption{Allowable fringe tracking bandwidth for varying stellar parameters. Overplotted in black is the LIFE catalog within 40\,pc, and bandwidth contours are shown in white. A baseline limit of 25-80\,m was imposed, and the OPD residuals were drawn from \cref{fig:FT_requirement}. Solid horizontal lines show the boundaries between stellar types of F, G, K and M; and dashed horizontal lines show the boundaries between LIFE's target stellar populations.}
    \label{fig:FT_bandwidth}
\end{figure*}

Finally, we combine it together to show the allowable fringe tracking bandwidth for the catalogue as a function of stellar parameters, whereby the residual is obtained from \cref{fig:FT_requirement}. In order to reduce the stability and internal metrology requirements on the mission, ideally the bandwidth is kept above 1\,Hz at a minimum, and is in the range of 10-100\,Hz ideally. It is clearly visible that solar-type stars allow a bandwidth of 10\,Hz or more, and that all targets from \added{stellar types F0-M3} allow a bandwidth of \added{at least} 1\,Hz. Using \cref{eq:mag}, for a bandwidth of 10\,Hz, and an OPD residual of 2\,nm RMS, we have a LIFE limiting magnitude of about $m_K=7.5$, which is fainter than nearly all stars of \added{type F0-K5} (solar-type stars) within 40\,pc. 

The very late M-dwarfs fare poorly, however, as most targets further than 10\,pc require more than 1\,s integration times. Nevertheless, the combination of these stars being of low scientific priority and the incompleteness of the catalogue at these effective temperatures indicates that this will likely not be a problem for the mission. Furthermore, if LIFE does require that these stars are able to be observed; if the control system requires a faster bandwidth; or if a lower fringe tracking residual is required, the spectral bandwidth could always be increased beyond the K-band to utilise more of the stellar light. 

\section{Conclusion}

As we have seen in this study, the effect that the baseline has on a nulling interferometer mission like LIFE is large and varied, making defining an optimal baseline per target, and baseline limits for the mission, challenging. Nevertheless, we have presented an \added{astrophysically} motivated \added{method} of estimating the baseline for both detection and characterisation, albeit with a strong caveat that it only maximises the bulk SNR integrated over the bandpass. One of the most important future avenues for study is the baseline strategy for LIFE during characterization. Prioritising the science cases and identifying what atmospheric species have the most importance based on initial spectral information will be critical \added{to decide which} baselines are optimal, and even whether LIFE should switch baselines half-way through an observation.

From a detection perspective, however, we found that the baseline range can be significantly reduced without losing much in the way of performance, \added{whether that be planet yield or} fringe tracking performance. In fact, \added{baseline limits affect} the visibility of the stellar sample minimally. A proposed reduction in the baseline limits of 25-80\,m results in a minor 5\% performance reduction, though it should be emphasised that this is merely one point in a trade space that should be further \added{examined} based on the ability to implement the mission (such as spacecraft size, formation flying margin and inter-spacecraft metrology precision). 

Discrete baselines also perform \added{modestly}, with a mere three baselines on the order of 30, 50 and 90\,m producing a $<$10\% drop in performance. However, we emphasise that part of the strength of LIFE is the flexibility to choose the baseline based on the target and \added{prioritised} spectral features. Giving up this advantage must be carefully weighed up against any perceived advantage in implementation.

Ultimately, the decision of what baselines LIFE can support, whether that be a continuous range or varied implementations of discrete baselines, is a key choice that affects many instrumental parameters and scientific performance. Further work in characterisation and wavelength-dependent optimisation is needed, and a full trade study with other architectural choices should be undertaken, before settling on the final implementation of the LIFE space mission.

\begin{acknowledgements}
      Part of this work has been carried out within the framework of the National Centre of Competence in Research PlanetS supported by the Swiss National Science Foundation under grants \texttt{51NF40\_182901} and \texttt{51NF40\_205606}. JTH, FAD, A.F. and SPQ acknowledge the financial support of the SNSF. R. L. has received funding from the Research Foundation - Flanders (FWO) under grant No. 1234224N. This project was supported by Rudolf Bär via the ETH Zurich Foundation. No AI tools were used in either the analysis or writing of this manuscript. The scripts and data used in this analysis can be found publicly at \doi{10.5281/zenodo.19366956}.
\end{acknowledgements}

\begin{contribution}
J.T.H contributed the concept, methods, analysis and writing of the manuscript. T.B contributed to the mathematical analysis, fringe tracking and FOV discussion. F.A.D assisted with yield simulations and interpretation. P.H assisted with baseline optimisation and interpretation. A.F and R.L assisted with the focus of the manuscript in the context of the LIFE instrumentation science community. L.S and J.K provided the star catalog and planet populations respectively. A.M.G and S.P.Q provided access to resources for the study. All authors commented on the manuscript.
\end{contribution}

\software{This article has made use of the following software packages: \textsc{LIFEsim} \citep{Dannert-2022-ID11}, 
          \textsc{Astropy} \citep{2013A&A...558A..33A,2018AJ....156..123A,2022ApJ...935..167A,2013A&A...558A..33A},
          \textsc{CMasher} \citep{vanderVelden2020},
          \textsc{cubature} \citep{castro_cubature_2025},
          \textsc{joblib} \citep{Joblib},
          \textsc{lmfit} \citep{lmfit},
          \textsc{Mathematica} \citep{Mathematica},
          \textsc{matplotlib} \citep{matplotlib},
          \textsc{MeanStars} \citep{savransky_meanstars},
          \textsc{multiprocess} \citep{multiprocess,multiprocess2}
          \textsc{NumPy} \citep{numpy},
          \textsc{pandas} \citep{reback2020pandas,mckinney-proc-scipy-2010},
          \textsc{PPop} \citep{Kammerer-2018-ID74,Kammerer_PPop_Code},
          \textsc{PyTables} \citep{pytables},
          \textsc{PyYAML} \citep{pyyaml}, and
          \textsc{SciPy} \citep{scipy}.
          }

%
\bibliographystyle{aasjournalv7} 
\bibliography{report} 

\appendix
\crefalias{section}{appendix}




\section{Pseudo-analytical derivations of flux sources}
\label{app_sources}

In this appendix, we re-derive the flux sources used in our simulations in as close to an analytic matter as possible, to allow for better understanding of the relationships between the parameters and to speed up calculations. We also include secondary effects including that of limb darkening. 

\subsection{Planet signal}

As most of LIFE's target planets are located in parameter spaces unlikely to be found by indirect techniques such as transits, astrometry and radial velocity, it is foreseen that LIFE will spend a non-negligible fraction of its mission time detecting the planets it will then stare at during a characterisation phase. The detection phase is thus defined by optimising the instrument to detect a \added{predetermined} type of planet at a certain distance from its host star, and then rotating the array to produce a modulation signal that should identify said planet (if it exists) with a given confidence within a certain amount of time. The process with which the signal is to be extracted and analysed, and the requirement to look multiple times for reasons including constraining the orbit of the planet is beyond the scope of this work. We direct the reader to papers such as \cite{Huber-2025-ID58} for more information.

Rotating the array to create a modulating signal can be thought of from the frame of reference of the array itself: as the array rotates, the planet simply follows a circular path through the interferometer's ``transmission map'', which is the response function for a given angular position on the sky. More precisely, for the double Bracewell, this the modulation of the planet through the ``differential'' transmission map \citep[see][for more details]{Dannert-2022-ID11,Hansen-2022-ID6}. The double Bracewell differential transmission function, normalised to a single telescope flux, is given by \cite{Dannert-2022-ID11}:
\begin{equation}
\label{eq:trans_diff}
    T_\text{diff}(B,\lambda,\theta,\phi) = \sin\left(\frac{\pi B \theta}{\lambda} \cos(\phi)\right)^2 \sin\left(\frac{2\pi q B \theta}{\lambda}\sin(\phi)\right),
\end{equation}
where $B$ is the baseline, $\lambda$ the wavelength, $q$ the imaging to nulling baseline aspect ratio (assumed to be 6:1 following \cite{Lay-2006-ID26}), and $\theta,\phi$ are the polar angular coordinates on the plane of the sky. 

For the purpose of this work, and in a similar vein to previous studies \citep{Dannert-2022-ID11,Hansen-2022-ID6}, we define a planet as being detected by obtaining a total SNR of 7 after an integer number of rotations. As such we define the ``modulation efficiency'', $\xi$, of a given angular separation as the root-mean-squared azimuthal average of the differential map at that separation \citep{Lay-2004-ID24,Dannert-2022-ID11,Hansen-2022-ID6}. This value denotes the fraction of the signal that is available for use in signal extraction. Mathematically, 
\begin{align}
    \xi(B,\lambda,\theta) = \sqrt{\frac{1}{2\pi}\int_0^{2\pi}T_\text{diff}(B,\lambda,\theta,\phi)^2d\phi}.
\end{align}
Solving this integral gives us an analytical expression for the modulation efficiency. Substituting $x = \frac{2\pi B\theta}{\lambda}$, and then letting $q=6$ as per the reference design:
\begin{align}
\label{eq:mod_eff}
        \xi(x) &= \frac{1}{4}\sqrt{3-4J_0\left(x\right)+J_0\left(2 x\right)-3J_0\left(2 qx\right)-J_0\left(2 x \sqrt{1+q^2}\right)+4J_0\left(x \sqrt{1+4q^2}\right)}\nonumber\\
        &= \frac{1}{4}\sqrt{3-4J_0\left(x\right)+J_0\left(2 x\right)-3J_0\left(12 x\right)-J_0\left(2 x \sqrt{37}\right)+4J_0\left(x \sqrt{145}\right)},
\end{align}
where $J_0$ is the zeroth order Bessel function of the first kind. 

\begin{figure}
    \centering
    \includegraphics[width=0.5\linewidth]{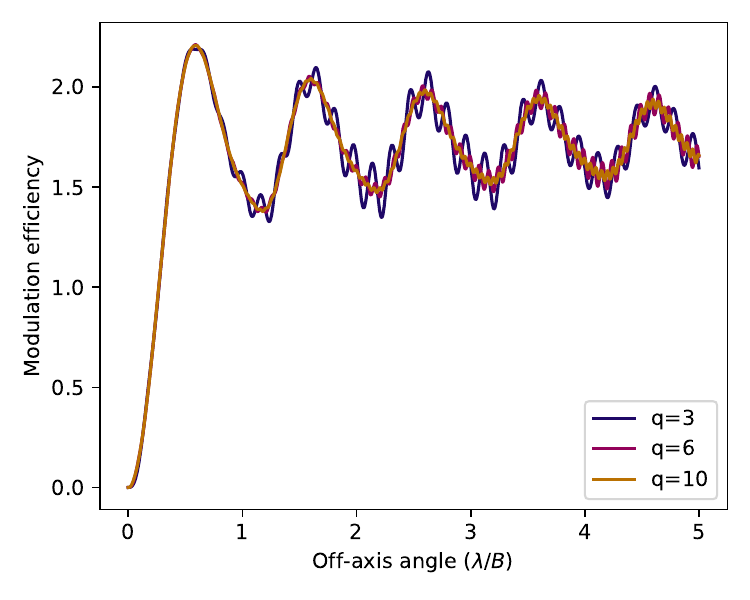}
    \caption{Modulation efficiency curve of the Double Bracewell for varying aspect ratios as a function of off-axis angle.}
    \label{fig:mod_eff}
\end{figure}

We plot the modulation efficiency curves as a function of off-axis angle (in units of $\lambda/B$) for a few aspect ratios ($q = 3,6,10$) in \cref{fig:mod_eff}. From here, we note that close in to the star (or for short baselines), the signal drops off as the null becomes bigger. For large separations, the signal oscillates, tending to an average transmission of $\sqrt{3}/4 = 0.433$ telescope fluxes. The amount of oscillation at large separations decreases for larger aspect ratios, as the number of fringe crossings increases and blurs out. The maximum, as has been noted in previous works \citep{Dannert-2022-ID11,Hansen-2022-ID6}, can be found at 0.59$\lambda/B$, and produces an average transmission of 0.55 telescope fluxes.

The total planet flux is then the product of the planet's spectral flux density $E_p$, the modulation efficiency, and the FOV taper function $\rho(D,\theta,\lambda)$, integrated over wavelength:
\begin{equation}
\label{eq:flux_p_det} 
    F_{p,det,i}(T_p,\delta_p,B,\theta,D) = \int_{\lambda_i}^{\lambda_{i}+\Delta\lambda
}E_p(T_p,\lambda,\delta_{p})\cdot\xi(B,\lambda,\theta)\cdot \rho(D,\theta,\lambda) d\lambda
\end{equation}
Here, $D$ is the aperture diameter, $T_p$ is the planet effective temperature and $\delta_p$ is the planet angular diameter. We typically will use a Gaussian function for the FOV, given in \cref{eq:fov} and discussed in \cref{app_FOV}. The angular separation of the planet, $\theta_p$ can be calculated from the semi-major axis and distance as $a/d$, and the angular diameter from the planet radius as $2R_p/d$. If using as a blackbody, we calculate the spectral flux density from the Planck function, or else substitute this in from an empirical spectra:
\begin{equation}
    \label{eq:sfd}
    E(T,\lambda,\delta) = \frac{\pi\delta^2}{4}B_\lambda(T, \lambda)
\end{equation}
Unfortunately, this integral cannot be solved analytically or via an approximation with a small amount of terms; instead, this must be computed numerically.

For the characterisation campaign, the overall concept is similar except that the position of the planet is known \textit{a priori}. As such, the array is optimised for the particular planet, rather than the planet archetype as in the search campaign. The other difference stems from a longer integration time and an emphasis of the SNR of various spectral lines, rather than the summed total signal to noise. This is strongly dependent on the science hypothesis being tested (such as the simultaneous detection of methane and carbon dioxide \citep{Thompson-2022-ID44}), and as these cases are still in the process of being developed and quantified, we focus the rest of our analysis on utilising the bulk SNR through weighting each spectral bin equally. The extension of this work to the retrieval of spectral features is nevertheless an important next step for the near future.

\added{We briefly note here that we are only considering single planets; many stellar systems will likely contain multiple sources in the field of view \citep[e.g.][]{Rowe_2014}. A question arises whether adding more planets degrades the ability to extract the signal of interest. This question is beyond the scope of this study, but earlier works indicate that signal extraction techniques can indeed recover multiple planets in a system \citep[e.g.][]{Dannert-2022-ID11,Matsuo-2023-ID5}. \cite{Cacaj-2025-ID73} also find that the probability of two planets being in close proximity to each other such that they pollute the measurement is relatively low ($<$3\%), though they also emphasise that being able to optimise for a given planet (i.e. having a range of long and short baselines) reduces the likelihood of this ``photobombing'' becoming problematic. Nevertheless, a future study that ties the choices of baseline and how they affect the distinguishing of multiple planetary sources is an important next step \citep[c.f.][]{Rutten-2026}}

\subsection{Stellar leakage}

Turning to stellar leakage, we need to identify a formulation for how much light is transmitted past the null due to the fact that the star is resolved. We will formulate this as the product of the star's radiance and a ``leakage'' factor ($\zeta$), all integrated over wavelength. We will not consider the aperture field of view here as the star will not be resolved enough to have off-axis coupling effects. We will also consider limb darkening, here approximated as a quadratic relation:
\begin{align}
\label{eq:limb_darkening2}
    LD(\theta) &= 1-a_1(1-\mu)-a_2(1-\mu)^2,\quad
    \mu = \sqrt{1-(2\theta/\delta_s)^2},\quad\theta\leq\delta_s/2,
\end{align}
where $\theta$ is the off-axis angle from the sky and $\delta_s$ is the stellar angular diameter. The parameters $a_1$ and $a_2$ are given through fitting to stellar measurements, such as those from \cite{Claret-2011-ID28}.

To calculate this leakage factor, we need to take an integral over solid angle of the product of the raw transmission map and the stellar limb; hence $\zeta$ has units of steradians. We will also include a visibility reduction parameter to incorporate the imperfection of the nulling instrument, related to the null depth by
\begin{equation}
    N_d = \frac{1-V}{1+V},
\end{equation}
whereby it follows from \cite{Birbacher-Hansen-2026} that the extra leakage results in a pertubated transmission of
\begin{equation}
    T'(\theta) = T(\theta)V + 0.5(1-V).
\end{equation}

Putting it all together gives
\begin{equation}
    \zeta(B,\lambda,\delta_s) = 0.5(1-V)+V\int_0^{2\pi}\int_0^{\delta_s/2}T_\text{raw}(B,\lambda,\theta,\phi)\cdot LD(\theta,\delta_s) \cdot\theta d\theta d\phi,
\end{equation}
with raw transmission map, akin to \cref{eq:trans_diff} given by \cite{Dannert-2022-ID11}:
\begin{align}
    T_\text{raw}(B,\lambda,\theta,\phi) = \sin\left(\frac{\pi B}{\lambda}\theta\cos(\phi)\right)^2\cos\left(\frac{\pi qB}{\lambda}\theta\sin(\phi)-\frac{\pi}{4}\right)^2.
\end{align}
Calculating the integral, we find that the leakage factor for $q=6$ is
\begin{equation}
    \zeta(x,\delta_s) = \frac{1}{8}k_0\pi\delta_s^2(2-V)-\frac{\pi V}{8\delta_sx^3}[k_1\delta_s^2x^2J_1(\delta_sx)+k_3\delta_sxJ_2(\delta_sx)-k_2(\delta_sx\cos(\delta_sx)-\sin(\delta_sx))],
\end{equation}
where $J_\nu$ are Bessel functions of the first kind, and we have made the substitutions:
\begin{align}
\label{eq:limb_params}
    k_0 &= \frac{1}{12}(6-2a_1-a_2) & k_2 &= a_1+2a_2 &
    k_1 &= 1-a_1-a_2 &
    k_3 &= -a_2 &
    x &= \frac{\pi B}{\lambda}.
\end{align}

Due to the very small angular diameters of the vast majority of main sequence stars ($\delta_s<1$\,mas), we take a Maclaurin series to obtain a very good approximation:
\begin{equation}
    \zeta(B,\lambda,\delta_{s}) \approx \frac{k_0\pi(1-V)\delta_s^2}{4} + \frac{15-7a_1-4a_2}{1920}\frac{V\pi^3B^2\delta_s^4}{\lambda^2} \approx \frac{15-7a_1-4a_2}{1920}\frac{\pi^3B^2\delta_s^4}{\lambda^2} \text{\quad [for $V\approx1$ ($N_d\approx 0$)]}.
\end{equation}
Hence we have shown the known result that for the double Bracewell, the stellar leakage is roughly proportional to the baseline squared. It is notable here to find that the flux due to stellar leakage is proportional to the fourth power of stellar angular diameter; hence the apparent stellar size has a massive impact on this noise term. 

As $\zeta$ already incorporates a solid angle dependence, the flux is obtained by simply integrating the product of the leakage with the Planck function over wavelength:
\begin{align}
\label{eq:flux_s} 
    F_{s,i}(T_s,\delta_s,B) &= \int_{\lambda_i}^{\lambda_{i}+\Delta\lambda
}B_\lambda(T_s,\lambda)\cdot\zeta(B,\lambda,\delta_{s})d\lambda.
\end{align}
We can approximate the dependence of limb darkening in terms of a scaling factor against the case of zero limb darkening. As described in \cref{app_limb_darkening}, for our specific range of stellar classes, we find that the stellar leakage flux with limb darkening is approximately a factor of 0.96 less than when neglecting limb darkening.

\subsection{Zodiacal and Exozodiacal dust}
Finally, we turn to the flux from the dust surrounding both our target system and around our Sun. The local zodiacal dust is the fundamental limit of our interferometer due to its diffuse nature; we cannot remove it via post-processing techniques. We model this via a very simple model from \cite{Glasse-2015-MIRI}, scaling a 270\,K blackbody emission by dust measurements fit to the celestial North pole \citep{Wright-1998-COBE} and with an additional factor of 1.2 for typical pointing variability. This emission is then multiplied by the integrated transmission map over the field of view:
\begin{align}
F_{z,i}(B,D) &= 
        4.3\cdot10^{-8}\int_{\lambda_i}^{\lambda_{i}+\Delta\lambda} \int_0^{2\pi}\int_0^\infty B_\lambda(270\,\text{K},\lambda)\cdot A(\theta,\lambda,D)\cdot T_\text{raw}(B,\lambda,\theta,\phi)\cdot\theta d\theta d\phi d\lambda\nonumber\\
&= \frac{4.3\cdot10^{-8}}{\pi D^2}(1-e^{-4B^2/D^2})\int_{\lambda_i}^{\lambda_{i}+\Delta\lambda}B_\lambda(270\,\text{K},\lambda)\lambda^2d\lambda
\approx \frac{4.3\cdot10^{-8}}{\pi D^2}\int_{\lambda_i}^{\lambda_{i}+\Delta\lambda}B_\lambda(270\,\text{K},\lambda)\lambda^2d\lambda,
\end{align}
where at the end we have assumed $B>>D$. Hence the photon count from the local zodiacal cloud is not dependant on any instrumental parameters other than the wavelength channel, as the diameter dependence cancels out. 

The exozodiacal dust will be modelled after \cite{Kennedy-2015-ID29}. We consider the dust in a face-on annulus around the star with a surface brightness profile given by
\begin{align}
    SB(r,z,L_\star,\lambda) = \Sigma_m(r,z,L_\star)\cdot B_\lambda(T_\text{disk}(r,L_\star),\lambda).
\end{align}
Here $\Sigma_m$ is the surface density of the disk, analogous to optical depth, and given by a power law:
\begin{align}
    \Sigma_m(r,z,L_\star) = z\Sigma_0\left(\frac{r}{\sqrt{L_\star/L_\odot}\cdot1\text{AU}}\right)^{-\alpha},
\end{align}
where $\Sigma_0$ is a scaling factor such that the parameter $z$ can be used for scaling the density in terms of ``zodis''; multiples of the density observed in our local zodiacal cloud. This value is $7.12\times 10^{-8}$ for a 1\,$z$ cloud at 1\,AU for a 1\,$L_\odot$ star. We adopt the power $\alpha = 0.34$ as fit to COBE observations in previous works \citep{Kelsall-1998-ID30,Kennedy-2015-ID29}. 

The exozodiacal emission is given by a Planck function with a radial temperature profile that scales with the host star luminosity:
\begin{align}
    T_\text{disk}(r,L_\star) = 278.3 \left(\frac{L_\star}{L_\odot}\right)^{0.25}\left(\frac{r}{1\text{AU}}\right)^{-0.5}.
\end{align}
Finally, we make the assumption following \cite{Kelsall-1998-ID30} and \cite{Kennedy-2015-ID29} that the disk emits between $r = 0.34$\,AU to 10\,AU.

We obtain the flux by taking the product of this surface brightness with the field of view function and the raw transmission map, and then integrating over solid angle and wavelength:
\begin{align}
    F_{e,i}(B,d,z,L_\star,D) &= \int_{\lambda_i}^{\lambda_{i}+\Delta\lambda}\int_0^{2\pi}\int_{0.034 \text{AU}/d}^{10 \text{AU}/d} SB(\theta d,z,L_\star,\lambda)\cdot A(D,\theta,\lambda) \cdot T_\text{raw}(B,\lambda,\theta,\phi) \cdot \theta d\theta d\phi d\lambda \\&= \frac{\pi}{2}\int_{\lambda_i}^{\lambda_{i}+\Delta\lambda}\int_{0.034 \text{AU}/d}^{10 \text{AU}/d} SB(\theta d,z,L_\star,\lambda)\cdot A(D,\theta,\lambda) \cdot \left(2-V-VJ_0\left(\frac{2\pi B\theta}{\lambda}\right)\right) \cdot \theta d\theta  d\lambda.
\end{align}
Unfortunately, this double integral is quite unwieldy for computations and does not appear to have a good analytical approximation. Nevertheless, with proper numerical optimisations, this is sufficient to undertake the following analysis in a reasonable amount of time.

\added{In this model, we have explicitly considered a face-on exozodiacal disk. The LIFE measurement scheme, using the double-Bracewell combiner, is able to remove any contributions from axisymmetric sources. Hence, even if the disk is inclined, the resultant photon flux is the same assuming symmetry is preserved. There are a number of ways this symmetry could be broken, including forward scattering, clumps, and disk offsets due to gravitational attraction \citep{Defrere-2010}. These structures may hinder the achievable SNR, as they will appear as a modulated signal in the measurement. However, assuming the nominal exozodical dust density published by the HOSTS survey \citep{Ertel2020}, these effects would still not be the limiting noise contribution for all but the most inclined disks \citep{Quanz-2022-ID17}. Observations to better constrain the exozodiacal density distribution, and to identify the importance of forward scattering of warm and hot exozodiacal dust between 4 and 10\,\textmu m, is critical to understand whether these assumptions are indeed well founded. Instruments such as the Large Binocular Telescope Interferometer (LBTI) and NOTT on the Very Large Telescope Interferometer (VLTI) plan to make progress on this front \citep[see e.g.][]{Defrere-2022,Ertel-2025}, and there is also ongoing work to better simulate the effects of these asymmetries on the LIFE measurement. We direct the reader to the studies of \cite{Defrere-2010} and \cite{Quanz-2022-ID17} for a more thorough discussion on these effects.}

\section{Discussion on limb darkening parameters}
\label{app_limb_darkening}
To determine which limb darkening parameters to use for both the stellar leakage and for fringe tracking, we first define a stellar subset consisting of the following parameter cuts:
\begin{enumerate}
    \item Stellar types of F through M (2500\,K to 7000\,K)
    \item $-1 \leq [FeH] \leq 1$, chosen due to our targets lying in the solar neighbourhood (see e.g. \cite{Haywood-2001-ID45})
    \item $\text{Log}(g) \geq 3.5$, derived from our stars being main sequence and the above stellar type range.
    \item Microturbulence below 2\,km/s, again from our stellar cut being relatively cool main sequence stars (see \cite{Steffen-2013-ID46})
\end{enumerate}
For the following analysis, we assume a parameterisation based on a quadratic relation as given in \cref{eq:limb_darkening2}:
\begin{align}
    LD(\theta) &= 1-a_1(1-\mu)-a_2(1-\mu)^2,\quad
    \mu = \sqrt{1-(2\theta/\delta_s)^2}\nonumber, \quad\theta\leq\delta_s/2.
\end{align}

\begin{figure}
    \centering
    \includegraphics[width=0.6\linewidth]{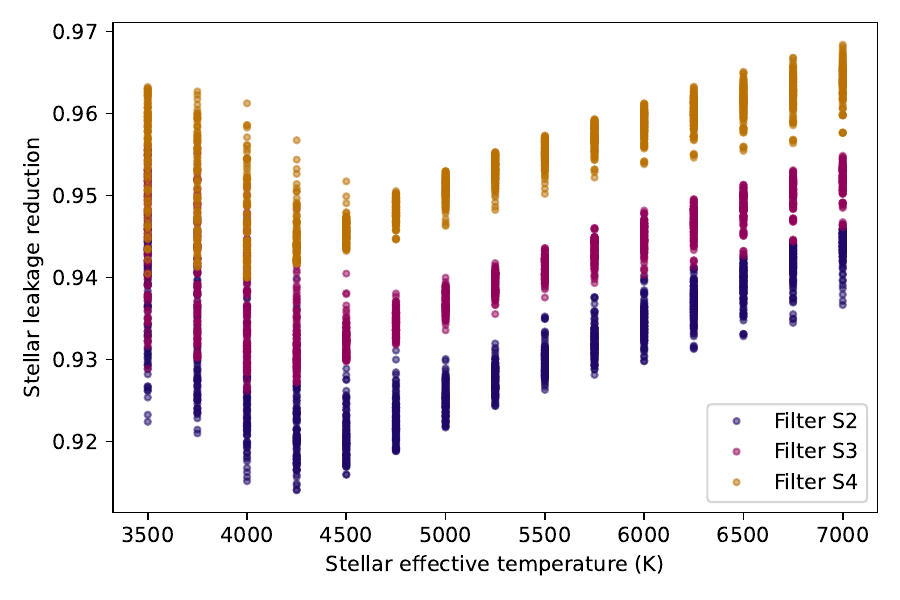}
    \caption{The reduction in stellar leakage due to limb darkening (\cref{eq:stellar_leakage_limb_darkening}) for various stellar models. Each point is an element of the table of parameters from \cite{Claret-2011-ID28}, assuming the parameter cuts discussed in \cref{app_limb_darkening}. The colours represent various Spitzer filters, with darker colours inferring shorter wavelengths.}
    \label{fig:limb_darkening_MIR}
\end{figure}

To start with the limb darkening dependence on stellar leakage, we can absorb both quadratic parameters into a single number describing the reduction of stellar leakage compared to the zero limb-darkening case, in the case of an instrument where the null depth $N_d\approx 0$:
\begin{equation}
\label{eq:stellar_leakage_limb_darkening}
    F_{s,i}(T_S,\delta_s,B) \approx \left(1-\frac{1}{15}(7a_1+4a_2)\right)\cdot F_{s,i}(T_S,\delta_s,B)_\text{$a_1$,$a_2=0$}.
\end{equation}
Using the table of limb darkening parameters from \cite{Claret-2011-ID28}, we then plot this factor across our stellar sample as described above in \cref{fig:limb_darkening_MIR}. The limb darkening reduction is plotted for the three Spitzer filters S2, S3 and S4 spanning roughly 4-5\,µm, 5-6.5\,µm and 6.5-9.5\,µm respectively. We can see that the reduction is modest and does not change substantially over the sample. To err on the conservative side and to simplify the calculations, we adopt a reduction factor of 0.96 for all stellar types and wavelengths.

\begin{figure}
    \centering
    \includegraphics[width=0.6\linewidth]{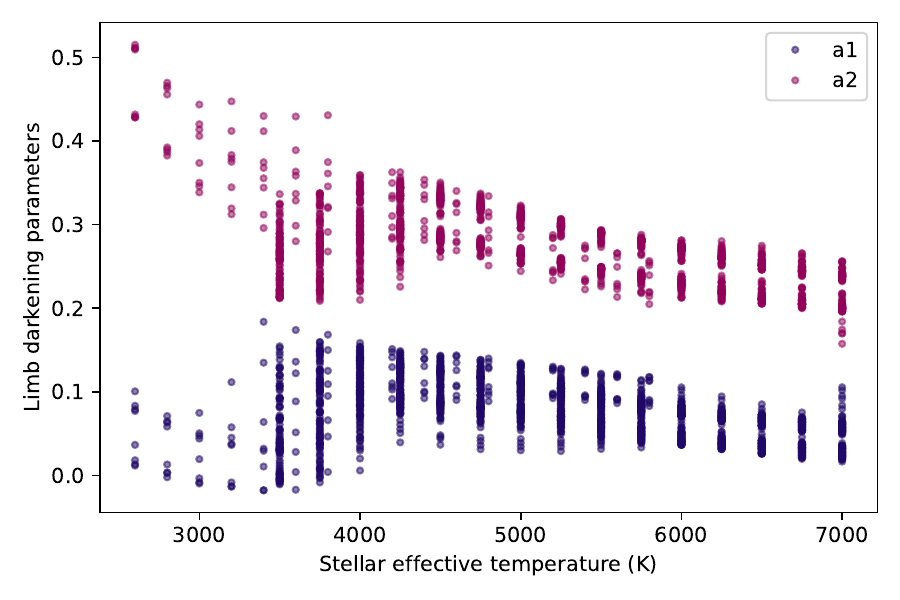}
    \caption{Limb darkening parameters in the Johnson K-band for various stellar models. Each point is an element of the table of parameters from \cite{Claret-2011-ID28}, assuming the parameter cuts discussed in \cref{app_limb_darkening}.}
    \label{fig:limb_darkening_FT}
\end{figure}

For fringe tracking, limb darkening will carry more of an effect due to the shorter wavelengths we plan on using (specifically the Johnson K-band), as well as the strong dependence it has on the visibility function (\cref{eq:visibility}). Utilising the same cut as described above, we plot the $a_1$ and $a_2$ limb darkening parameters for the K-band filter in \cref{fig:limb_darkening_FT}. While there is some structure here, overall it is relatively constant over wavelength, and so we adopt a constant values for our parameters of $a_{1,\text{K}} = 0.076$ and $a_{2,\text{K}} = 0.27$. This results in a maximum squared visibility for an unresolved star of $V^2 = 0.83$.

\section{Effect of off-axis throughput}
\label{app_FOV}

The field of view (FOV) of the LIFE mission is directly tied to its current use of a single-mode spatial filter. The spatial filter has many advantageous effects, including greatly loosening the requirements on wavefront control and allowing the mission to reach sensitivity despite the local zodiacal background. Nevertheless, this also restricts the coupling efficiency - the amount of light that can be injected into the spatial filter - for sources that are off-axis, such as a planet, and thus sets the so-called ``outer working angle'', or OWA. This is defined as the angle whereby the throughput drops to 50\% of the on axis value. 

The coupling efficiency in terms of intensity for a circular aperture has already been calculated in the literature by \cite{Shaklan-1988-ID21,Guyon-2002-ID20}. Here, the fundamental mode of the single-mode fibre is approximated as a Gaussian, which deviates from the point spread function of a circular aperture (an Airy function), leading to a maximum coupling on-axis of 82\%. Assuming that the instrument is focused to maximise the coupling efficiency, we find that the coupling of the fibre is given by \cite{Guyon-2002-ID20}:
\begin{equation}
       \rho_\text{airy}(\theta,\lambda,D) = 8e^{-3.923\left(\frac{\theta D}{\lambda}\right)^2}\left(\int_0^\infty e^{-r^2}I_0\left(2.802\frac{\theta Dr}{\lambda}\right)J_1\left(\frac{\pi r}{1.402}\right)dr\right)^2,
\end{equation}
where $\theta$ is the off-axis angle, $\lambda$ the wavelength, $D$ the aperture diameter, $I_0$ the zeroth order modified Bessel function of the first kind, and $J_1$ the first order Bessel function of the first kind. 

An alternative approach is to apodise the beam into \added{a shape} that better approximates a Gaussian , which allows for up to an $\sim$ 18\% improvement in maximum coupling (allowing for 100\% on axis) and is currently considered in the LIFE reference design through a set of phase-induced amplitude apodisation (PIAA) optics \citep{Jovanovic-2017-ID18,Glauser-2024-ID1}. Such optics apodise the beam into a prolate-spheroid, which very closely resembles a Gaussian but with sharper wings (due to a finite pupil size) \citep{Jovanovic-2014-ID22}, \added{with minimal throughput loss}. The off-axis coupling of a Gaussian beam into a single mode fibre is derived in Appendix A.4 of \cite{Birbacher-Hansen-2026}, and is given by:
\begin{equation}
    \rho_\text{gauss}(\theta,\lambda,D) = e^{-\frac{\pi^2}{4}\left(\frac{\theta D}{\lambda}\right)^2},
\end{equation}
which is similar to the first factor of the Airy function coupling equation.

\begin{figure}
    \centering
    \includegraphics[width=0.9\linewidth]{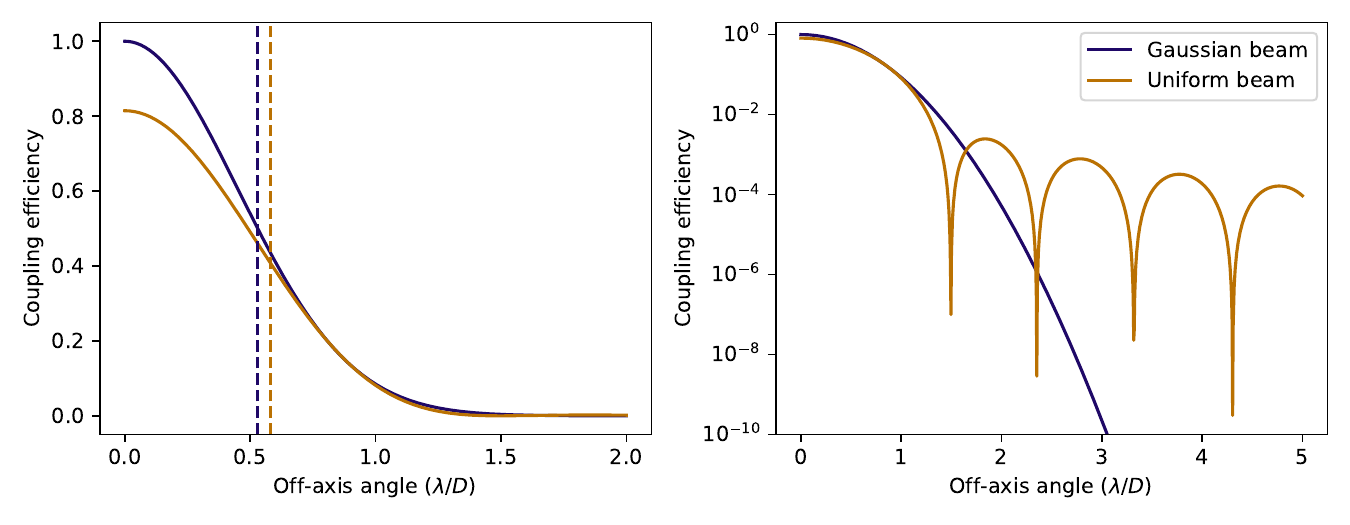}
    \caption{Field of view coupling efficiency for a Gaussian and a uniform beam as a function of off-axis angle. Dashed lines indicate the outer working angle (50\% of the maximum coupling efficiency). \textit{Left:} linear scaling; \textit{Right:} logarithmic scaling.}
    \label{fig:FOV_function}
\end{figure}

We plot the two functions in terms of $\lambda/D$ in \cref{fig:FOV_function}, where it is apparent that other than the decreased maximum coupling, the functions are similar. The OWA for the Gaussian beam is 0.53 $\lambda/D$, whereas the uniform beam, having wings in the Airy function, is slightly higher at 0.58 $\lambda/D$; \added{hence the latter couples better beyond 1.5$\lambda/D$}. We also see that the FOV is determined by purely the diameter of the aperture and the wavelength of light, with a larger diameter indicating a smaller field of view. Going forward, we will use the Gaussian form of the FOV (i.e. $\rho = \rho_\text{gauss}$), both because of its simpler functional form, and because PIAA optics are currently assumed for the LIFE instrument. 

\begin{figure}
    \centering
    \includegraphics[width=0.7\linewidth]{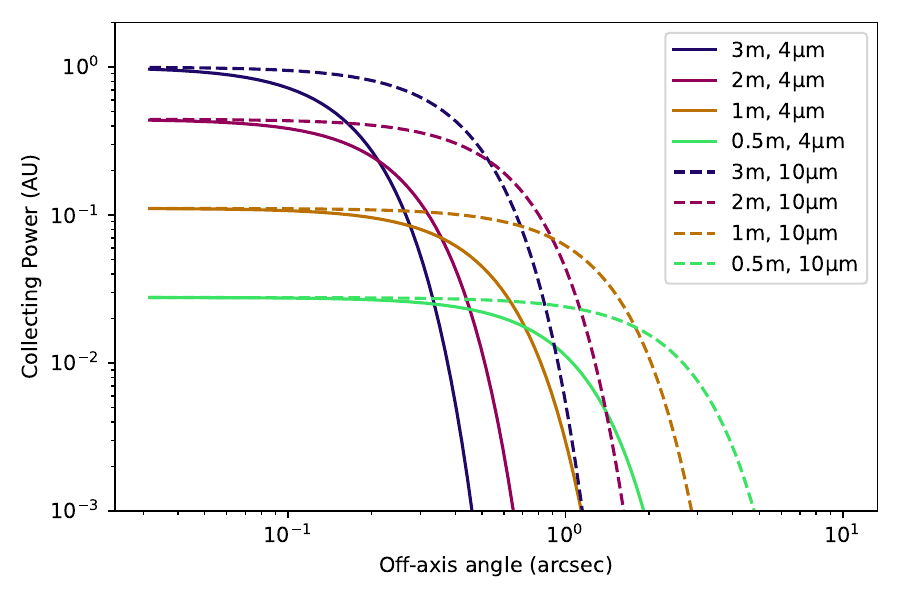}
    \caption{Collecting power of different telescope sizes as a function of off-axis angle, arbitrarily scaled such that an on-axis source for a 3\,m mirror is one. Shown for wavelengths of 4\,µm (left) and 10\,µm (right). Equivalent to the coupling efficiency plot in \cref{fig:FOV_function} multiplied by the collecting area of the primary mirror.}
    \label{fig:FOV_collection_area}
\end{figure}

A question arises as to whether it is advantageous to reduce (or stop-down) the aperture such that the increase in off-axis coupling efficiency overcomes the reduction in collecting area.  In \cref{fig:FOV_collection_area}, we plot the coupling efficiency multiplied by the collecting area of the collector for a number of diameter sizes and wavelengths. We can see that indeed the product of these two effects lead to situations for off-axis sources to be detected more efficiently with smaller apertures than larger ones.

\begin{figure}
    \centering
    \includegraphics[width=0.85\linewidth]{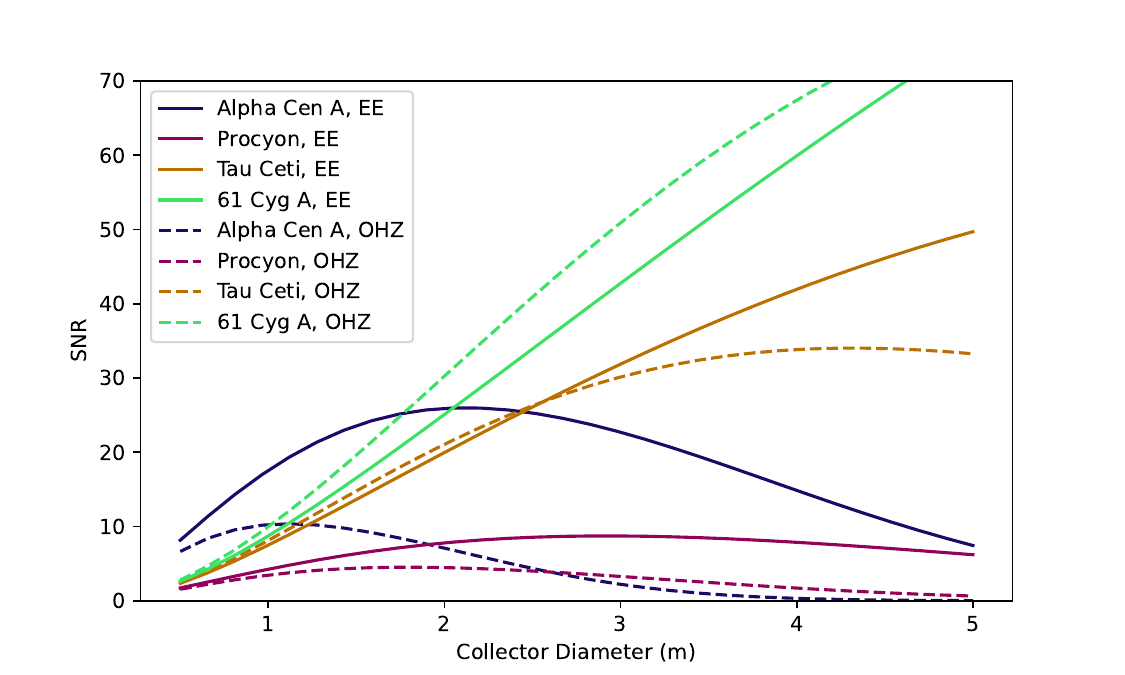}
    \caption{SNR for a variety of planet systems as a function of aperture diameter. EE refers to an ExoEarth twin at 1\,AU, scaled by the square root of the stellar luminosity, and OHZ refers to an Earth placed at the outer edge of the habitable zone.}
    \label{fig:FOV_SNR}
\end{figure}

To investigate further, we look at the SNR for a number of fiducial cases as the FOV changes due to a varying diameter. We consider telescope dishes with diameters from 0.5 to 5\,m, and we analyse how the signal changes between four stellar systems:  $\alpha$ Cen A, Procyon, $\tau$ Ceti and 61 Cyg A. These four targets were chosen due to being among the closest solar-type (FGK) stars, and thus having the largest angular separation between star and habitable zone. Two different planet positions were used: an ``exoEarth'' at 1\,AU, scaled by the square root of the stellar luminosity; and an Earth at the outer edge of the habitable zone as defined by \cite{Kaltenegger-2017-ID48}. For the planets we use an Earth-equivalent spectrum \added{created via the Planetary Spectrum Generator (PSG, \cite{Villanueva-2018-ID72})} and a wavelength range spanning 4 to 18.5 microns, using the tools described in \cref{app_sources}. The results are plotted in \cref{fig:FOV_SNR}.

This investigation shows that for even very close-by G-star systems such as $\tau$ Ceti, one still obtains an increase in SNR with increasing diameter up to about 5\,m. Procyon does encounter some benefit to reducing the aperture if we assume a reference aperture size of 3-4\,m, although the benefit is marginal. There is one major exception to the above: $\alpha$ Cen A (and B). For this system, a prospective Earth-like planet would be outside the outer working angle and would have a significant throughput loss. Stopping down to a 2\,m or smaller aperture matters, as we get an integrated signal 50\% larger than the 3.5\,m case. Planets further out fare even worse. Essentially, if $\alpha$ Cen is to be a target, adding the ability to reduce the aperture size (or equivalently increasing the FOV) would lead to substantial SNR improvements. This leads to the question: for a single target, is it worth adding the instrumental complexity, especially if prospected candidates can be detected or characterised with other instruments in a more optimal capacity (e.g. TOLIMAN \citep{Tuthill-2024-ID66} or ELT/METIS \citep{LeCoroller-2022-ID65})? It is sensible to design the mission towards a statistical sample rather than the exception. Nevertheless, we identify that it can still be reached at the cost of an extra cryogenic mechanism adding a pupil stop.

We finish with a number of comments. Firstly, the above signal analysis considered the full wavelength range as would be the case in the detection campaign. For characterisation, shorter wavelengths may be desired for their particular diagnosis potential, and this increases the FOV impact. However, even when considering the flux solely between 4 and 8\,µm, an planet at the edge of the HZ around $\tau$ Ceti still does not benefit from stopping down the pupil for diameters less than 3.5\,m. 

Secondly, there are a few potential observation strategies to mitigate this problem entirely: namely that of either offsetting the fringe tracker from the centre of the field of view, similar to that of VLTI/GRAVITY \citep{Lacour-2019-ID40}, or by adjusting the angle of the fibre such that the planet is on-axis. For the former, rather than inserting the star at the centre of the FOV, the star is offset along with a predetermined phase delay to ensure the star location is nulled. The planet (or prospective planet) location is then placed at the centre, and to account for an unknown position angle, multiple exposures with varying offset angles may need to be provided to ensure detection. This strategy comes with many complications, including calculation of the fringe tracker offset and impact on the array modulation (namely, that it would no longer be symmetric about the optical axis). If a planet is known from a prior detection, however, then a simple tilting of the fibre may be sufficient to remove the FOV-induced throughput loss. Such strategies are perhaps the best way to consider observing a target like $\alpha$ Cen or any other planets at very large angular separations from the host star. 
For some planetary orbit inclination angles, this problem could even be negated by simply waiting for the planet to be viewed at smaller angular separations. As the orbital periods are on the order of a few years, this may not be a substantial wait.

Finally, the FOV is also limited by the spectral resolution ($R$) of the spectrograph. With the basic criterion of FOV $<\lambda R/B$, to ensure that the collected flux within a spectral channel has coherence (and can be interpreted), this leads to $R_\text{min} \approx B/D$, where $B$ is the largest baseline. So having large $D$ can help limit the requirements on the spectrograph. 

\section{Effects of stellar and planetary parameters on the optimal baseline}
\label{app_optimised}

\begin{figure}
    \centering
    \includegraphics[width=0.64\linewidth]{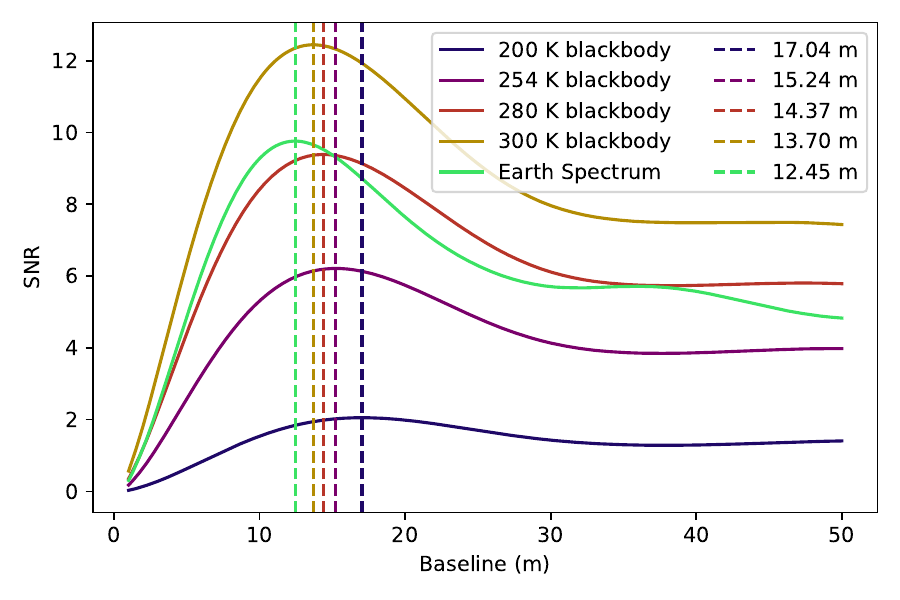}
    \caption{SNR as a function of baseline for various planet archetypes around a solar analogue star at 10\,pc. Vertical dashed lines represent the optimal baseline for each case.}
    \label{fig:optimised_plot_planets}
\end{figure}

Here, we describe and elaborate on the myriad of effects that can manipulate the choice of ``optimal baseline'', beyond that of the simple scaling law of \cref{eq:ref_wav}. First, in \cref{fig:optimised_plot_planets}, we plot the SNR as a function of baseline for five different planetary spectra. All planets have a radius of 1\,$R_\oplus$ and are located at 1\,AU from a Solar analogue at 10\,pc. Thus, the only change here is the planetary temperature and/or spectral shape - this in turn will only affect the planet signal component of the SNR calculation (\cref{eq:snr}) and not the noise. The vertical lines denote the optimal baseline for each of these planets. We can easily see that when comparing the temperatures of different blackbody sources, hotter planets have higher signal (due to a higher flux) and also shifts towards shorter baselines. This latter effect stems from having more signal at shorter wavelengths, which are better optimised at shorter baselines. 

Perhaps more interesting is the comparison between a blackbody with the effective temperature of Earth, and that of a realistic Earth spectrum \added{(created via PSG, \cite{Villanueva-2018-ID72})}. We can clearly see that the realistic Earth has a shift towards shorter baselines than a blackbody at higher temperatures. This is due to Earth's equivalent temperature being masked by absorption features at longer wavelengths, hence the true spectrum has higher signal at shorter wavelengths than a blackbody approximation would suggest.

\begin{figure}
    \centering
    \includegraphics[width=0.64\linewidth]{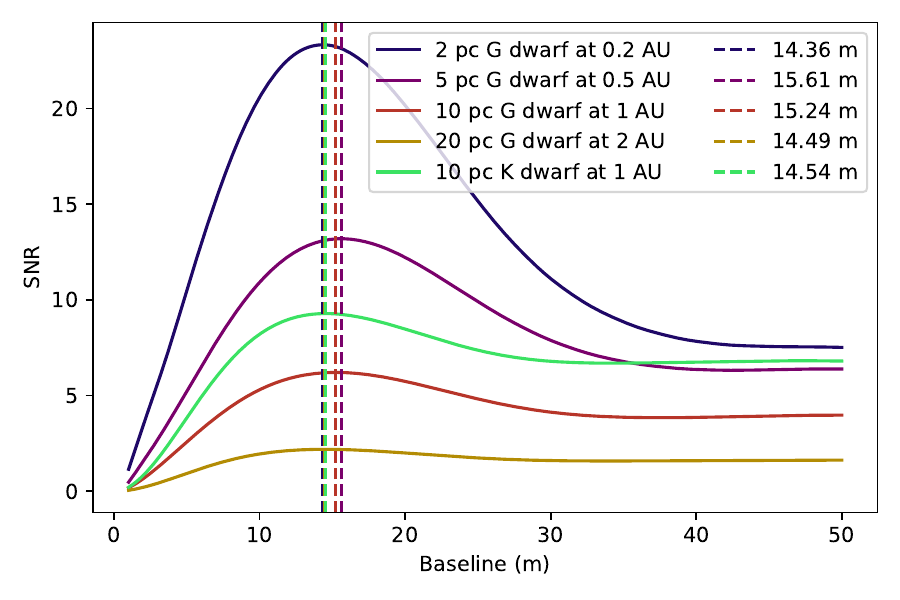}
    \caption{SNR as a function of baseline for different star archetypes considering an Earth analogue at 100\,mas from the star. The first four lines are for a G2 dwarf at various distances (but holding angular separation constant) and the latter a cooler K7 dwarf at 10\,pc. Vertical dashed lines represent the optimal baseline for each case.}
    \label{fig:optimised_plot_stars}
\end{figure}

Secondly, we now turn to modifying the stellar host rather than the planet. In \cref{fig:optimised_plot_stars}, we again plot the SNR as a function of baseline, but instead hold the planetary properties constant (as an Earth analogue). The angular separation is also held constant at 100\,mas: a planet's physical separation is changed proportional to the distance to the system. We plot four G2V Solar analogues at varying distances, as well as one cooler K7V star for comparison. There are a few noticeable, albeit minor, effects here. First, from 5 to 20\,pc, we see that the optimal baseline decreases. This stems from having less stellar leakage (the stellar disk is smaller) and so shorter wavelengths have a higher SNR, pushing the baselines shorter. In a similar vein, the cooler K-dwarf has less stellar leakage than its G counterpart and so pushes the baseline shorter to make use of the higher short wavelength signal. Perhaps surprisingly though, the 2\,pc star goes against this trend, pushing shorter despite being closer than the other stars. Here there is a second effect that begins to dominate: the stellar disk in this case is so large, and the leakage so high, that the optimiser would rather increase the size of the null through a shorter baseline than optimise for a longer wavelength with larger baselines. This emphasises the complexity of the noise sources impacting the choice of baseline.

Finally, we look at the effect of a planet around the same nearby (2\,pc) solar analogue, but at different separations. To look at second-order effects, we multiply the baseline by the separation, which would otherwise be the dominant effect. As we look for planets that are at increasing separations, we see that both the SNR gets larger and the baseline gets larger, as we can better optimise for separating the planet from the star. Past a certain point, however, we start to see the effects of the limited FOV come into effect (hence why the star is so close in this example). The baseline shifts further out as short wavelengths are reduced in throughput due to the FOV, and the overall SNR drops. 

\begin{figure}
    \centering
    \includegraphics[width=0.64\linewidth]{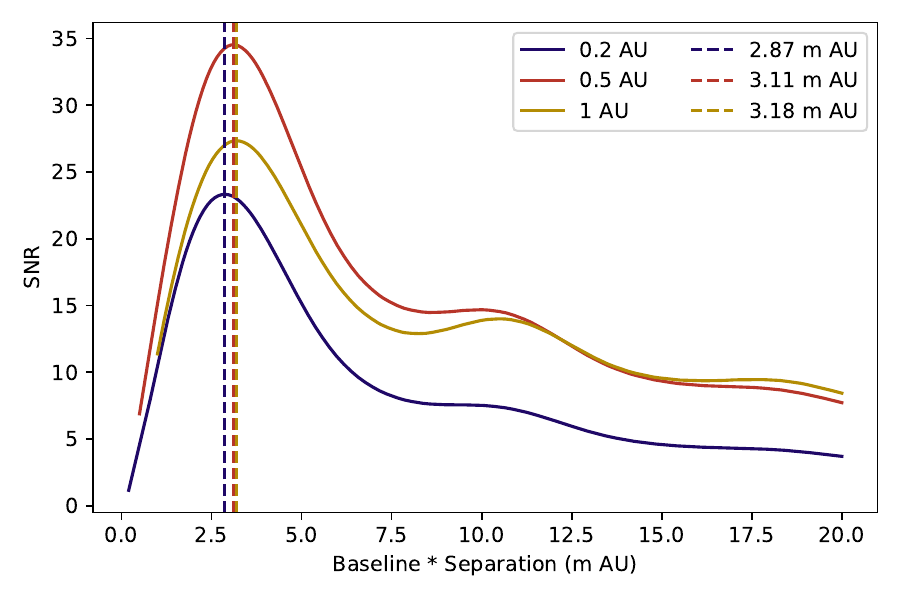}
    \caption{SNR as a function of baseline multiplied by separation for an Earth twin around a solar analogue at 2\,pc, shown for planets of varying separations. Vertical dashed lines represent the optimal baseline for each case.}
    \label{fig:optimised_plot_stars_FOV}
\end{figure}

\section{Baseline optimisation coefficients, residuals and yield comparison}
\label{app_coeffs}
\begin{figure}
    \centering
    \includegraphics[width=\linewidth]{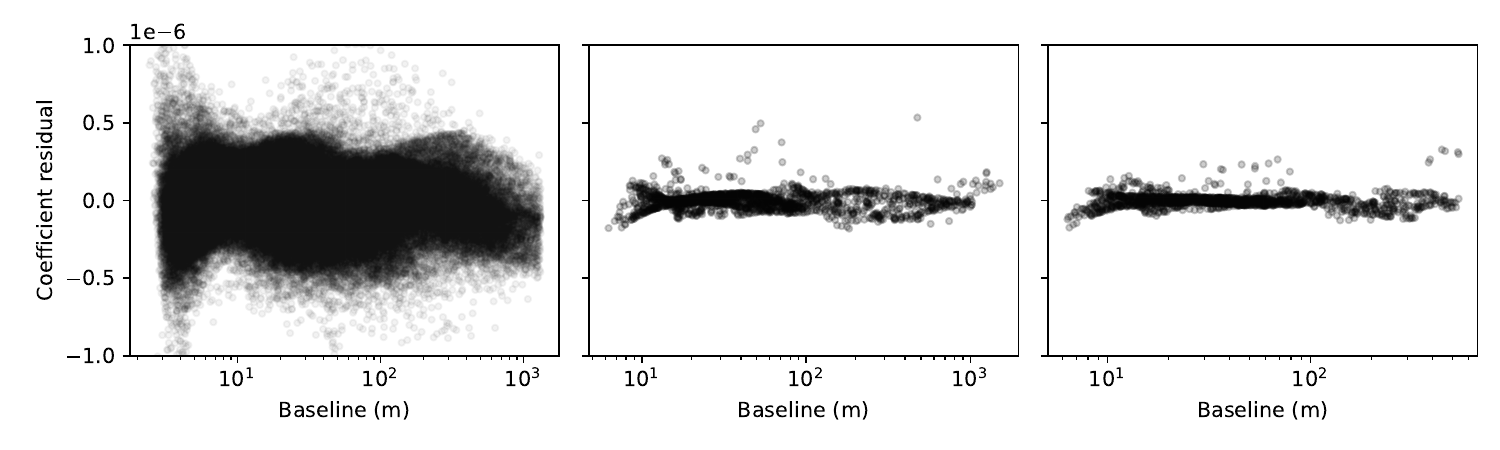}
    \caption{Residuals for each of the various MC parametric fits described in \cref{tab:opt_baseline_MC} as a function of baseline. Left: Characterisation model, with residual $\sigma$ of $1.7\times10^{-7}$\,m\,rad, Middle: \textit{Kepler} detection model, with residual $\sigma$ of $5.9\times10^{-8}$\,m\,rad, Right: Uniform detection model, with residual $\sigma$ of $4.4\times10^{-8}$\,m\,rad. The two detection models were run with fewer samples due to computational constraints.}
    \label{fig:MC_residuals}
\end{figure}

\begin{figure}
    \centering
    \includegraphics[width=0.8\linewidth]{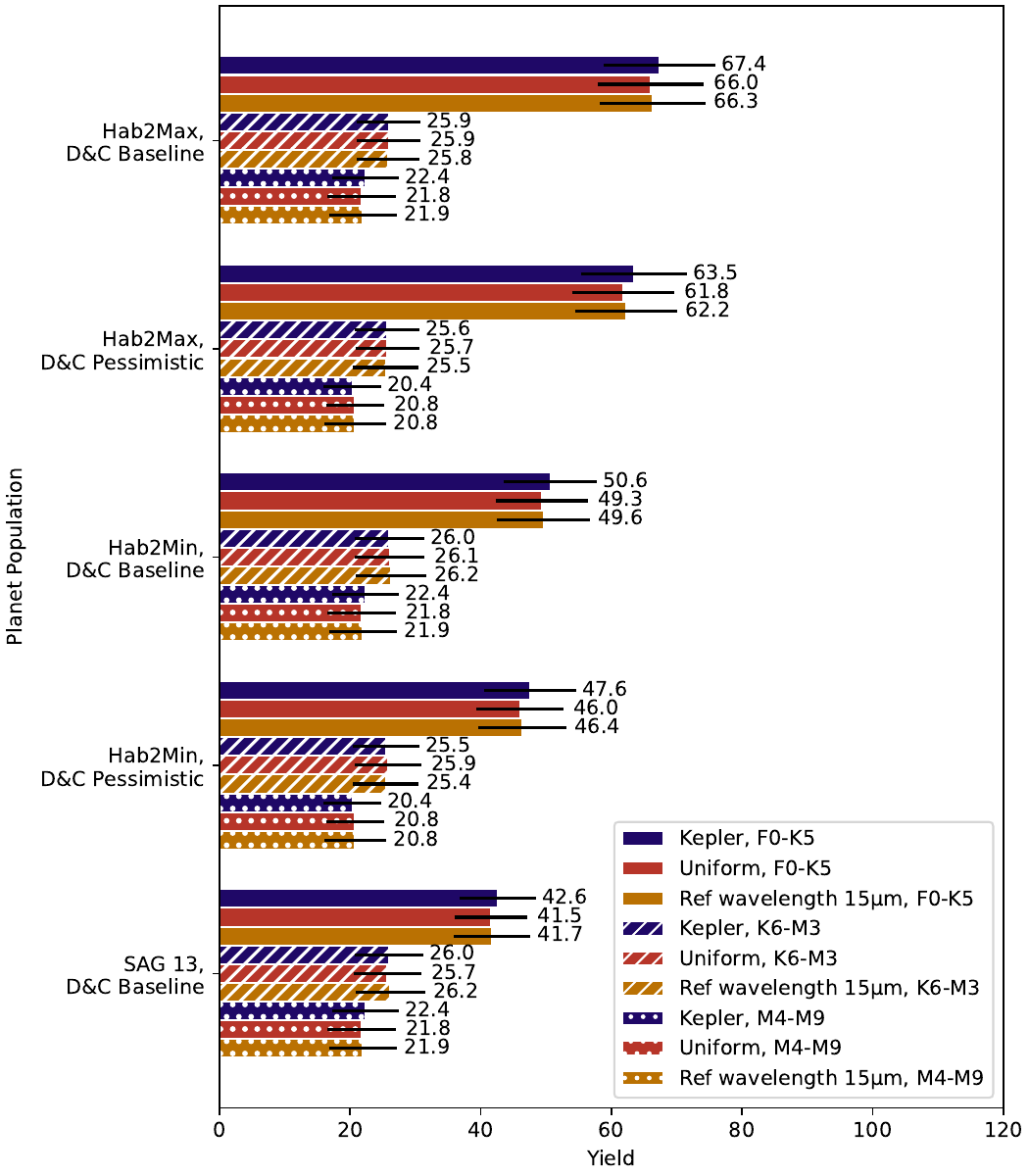}
    \caption{Number of exoplanet detections within the habitable zone for different baseline optimisations, stellar populations and statistical planet populations. The yields were run via \textsc{LIFEsim} \citep{Dannert-2022-ID11}, with modifications described in the text. The various optimisation routines (shown in different colours) are: setting the ``reference wavelength`` to 15\,µm as previously done in LIFE yield papers; optimizing assuming a semi-major axis distribution based on \textit{Kepler} statistics as described in the text; and optimizing assuming a uniform semi-major axis distribution. For the latter two, orbital projection effects were included, and the coefficients of the polynomial model for the baseline scaling can be found in \cref{tab:opt_baseline_MC}. The difference between solid, hatched and spotted bars represent different stellar populations described in the text, and the y-axis shows various planet populations: \textsc{hab2min} and \textsc{hab2max} for FGK stars from \cite{Bryson-2021-ID42}, and \textsc{SAG13} baseline model from \cite{Kopparapu-2018-ID55} respectively. M stars are drawn from \cite{Dressing-2015-ID69} statistics using either the baseline or pessimistic values as described. The mean detection yield is listed next to each bar and the standard deviation is shown via the black error bars.}
    \label{fig:opt_yields}
\end{figure}

We show the coefficients for the baseline optimisation polynomial (\cref{eq:k}) discussed in \cref{sec:optimisation} in \cref{tab:opt_baseline_MC}. We also show the residuals of the MC fits in \cref{fig:MC_residuals}. Finally, we show the raw yields of the \added{model verification simulation (\cref{sec:verification})} in \cref{fig:opt_yields}. We emphasise that the main purpose of this study is to perform a relative analysis as to how the performance changes with baseline length, or in the case of this plot, baseline optimisation model. We avoid drawing insight from the raw yields, as they are highly dependent on the assumptions made in the simulation (listed in \cref{sec:verification}), which can change substantially.

\begin{splitdeluxetable*}{lll|llllBlllllll}
\tablewidth{0pt} 
\tablecaption{Coefficients for the optimised baseline scale factor polynomial, shown in \cref{eq:k} \label{tab:opt_baseline_MC}}
\tablehead{
\colhead{Model} & \colhead{Stellar Subset}&  \colhead{Planet Temperature} &\colhead{$c_1$ ($T_p$)} & \colhead{$c_2$ ($T_p^2$)} & \colhead{$c_3$ ($a_\text{proj}$)} & \colhead{$c_4$ ($a_\text{proj}^2$)} & \colhead{$c_5$ ($L_\star^{0.5}$)} & \colhead{$c_6$ ($L_\star$)}& \colhead{$c_7$ ($d$)} & \colhead{$c_8$ ($d^2$)} & \colhead{$c_9$ ($L_\star^{0.5}d$)} & \colhead{$c_{10}$ ($\theta_p^2$)} & \colhead{$c_{11}$}
}
\startdata
\multirow{4}{*}{Characterisation} & FGK stars & $T_p > $ 350\,K & $-2.65\cdot10^{-8}$ & $2.00\cdot10^{-11}$ & $-7.87\cdot10^{-7}$ & $9.40\cdot10^{-8}$ & $1.18\cdot10^{-6}$ & $-2.26\cdot10^{-7}$ & $-7.32\cdot10^{-8}$ & $1.20\cdot10^{-9}$ & -- & $5.16\cdot10^{5}$ & $1.29\cdot10^{-5}$ \\
 & M Stars & $ T_p > $ 350\,K & $-2.64\cdot10^{-8}$ & $2.04\cdot10^{-11}$ & $-5.35\cdot10^{-6}$ & $4.67\cdot10^{-6}$ & $1.44\cdot10^{-5}$ & $-2.72\cdot10^{-5}$ & $-1.03\cdot10^{-7}$ & $2.05\cdot10^{-9}$ & -- & $6.00\cdot10^{5}$ & $1.23\cdot10^{-5}$ \\
 & FGK stars & $T_p \leq$ 350\,K & $-1.90\cdot10^{-8}$ & $1.16\cdot10^{-12}$ & $-3.88\cdot10^{-7}$ & $4.08\cdot10^{-8}$ & $7.32\cdot10^{-7}$ & $-1.27\cdot10^{-7}$ & $-2.26\cdot10^{-8}$ & $2.88\cdot10^{-10}$ & -- & $1.61\cdot10^{5}$ & $1.21\cdot10^{-5}$ \\
 & M stars & $T_p \leq$ 350\,K & $-1.85\cdot10^{-8}$ & $-5.06\cdot10^{-12}$ & $-1.86\cdot10^{-6}$ & $1.58\cdot10^{-6}$ & $5.60\cdot10^{-6}$ & $-1.06\cdot10^{-5}$ & $-3.31\cdot10^{-8}$ & $5.46\cdot10^{-10}$ & -- & $1.72\cdot10^{5}$ & $1.22\cdot10^{-5}$ \\ \midrule
\multirow{2}{*}{\begin{tabular}[c]{@{}l@{}}Detection:\\ \textit{Kepler} distribution\end{tabular}} & FGK stars & -- & -- & -- & -- & -- & $-2.57\cdot10^{-7}$ & $2.07\cdot10^{-7}$ & $4.87\cdot10^{-8}$ & $-4.09\cdot10^{-10}$ & $-2.16\cdot10^{-8}$ & -- & $9.34\cdot10^{-6}$\\
 & M stars & -- & -- & -- & -- & -- & $-7.41\cdot10^{-6}$ & $1.43\cdot10^{-5}$ & $1.19\cdot10^{-7}$ & $-2.31\cdot10^{-9}$ & $-6.94\cdot10^{-9}$ & -- & $9.71\cdot10^{-6}$\\ \midrule
\multirow{2}{*}{\begin{tabular}[c]{@{}l@{}}Detection:\\ Uniform distribution\end{tabular}} & FGK stars & -- & -- & -- & -- & -- & $1.99\cdot10^{-7}$ & $6.45\cdot10^{-8}$ & $2.61\cdot10^{-8}$ & $-8.18\cdot10^{-11}$ & $-1.70\cdot10^{-8}$ & -- & $8.30\cdot10^{-6}$ \\
 & M stars & -- & -- & -- & -- & -- & $-2.92\cdot10^{-6}$ & $4.45\cdot10^{-6}$ & $6.59\cdot10^{-8}$ & $-1.46\cdot10^{-9}$ & $2.85\cdot10^{-8}$ & -- & $8.43\cdot10^{-6}$  \\
\enddata
\tablecomments{The polynomial is derived piecewise for different planet temperatures ($T_p$) and stellar types. The boundary between M and K type stars is set at 3900\,K. There are also three different models used. ``Characterisation'' assumes maximising SNR for a single target, whereas ``Detection'' aims to achieve the maximum number of detections within the habitable zone. The latter is also divided between using a semi-major axis distribution based on \textit{Kepler} statistics or assuming a uniform distribution in semi-major axis.}
\end{splitdeluxetable*}

\end{document}